\UseRawInputEncoding
\documentclass[11pt, reqno]{amsart}
\usepackage{amsfonts,latexsym, enumerate}
\usepackage{amsmath}
\usepackage{amscd}
\usepackage{float,amsmath,amssymb,mathrsfs,bm,multirow,graphics}
\usepackage[dvips]{graphicx}
\usepackage[percent]{overpic}
\usepackage{amsaddr}
\usepackage{todonotes}
\usepackage[numbers,sort&compress]{natbib}
\usepackage{grffile}

\numberwithin{equation}{section}
\newcommand{\R}{{\mathbb R}}

\newcommand{\C}{{\mathbb C}}

\newcommand{\Z}{{\mathbb Z}}

\DeclareMathOperator{\Tr}{Tr}

\DeclareMathOperator{\Ln}{Ln}

\newcommand{\BCS}{\text{\upshape BCS}}
\newcommand{\half}{\text{\upshape half}}
\newcommand{\AF}{\text{\upshape AF}}
\newcommand{\F}{\text{\upshape F}}
\newcommand{\BI}{\text{\upshape P}}

\newtheorem{theorem}{Theorem}[section]

\newtheorem{lemma}[theorem]{Lemma}

\newtheorem{remark}[theorem]{Remark}

\newtheorem{figuretext}{Figure}

\usepackage[colorlinks=true]{hyperref}
\hypersetup{urlcolor=blue, citecolor=red, linkcolor=blue}

\input epsf
\title[Hartree--Fock analysis of the bilayer Hubbard model]{Hartree--Fock analysis of the bilayer Hubbard model}

\author{E. Langmann$^{1}$ and J. Lenells$^{2}$}
\address{$^1$Department of Physics, KTH Royal Institute of Technology, \\ 106 91 Stockholm, Sweden
	\\
$^{2}$Department of Mathematics, KTH Royal Institute of Technology, \\ 100 44 Stockholm, Sweden\\
}
\email{langmann@kth.se}
\email{jlenells@kth.se}
    
\begin{document}
\begin{abstract}
We study the half-filled Hubbard model on two parallel square lattices with on-site repulsion $U>0$ and with nearest-neighbor hopping specified by a constant $t>0$ within each plane and by another constant $t_\perp \geq 0$ between the planes. Working in the Hartree--Fock approximation, we obtain asymptotic formulas for the N\'eel temperature and the antiferromagnetic gap ratio in the weak coupling limit. We also determine the behavior of the antiferromagnetic gap ratio near the critical curve separating the paramagnetic 
and antiferromagnetic regions in the phase diagram. 
Furthermore, we present numerical phase diagrams away from half-filling, and we make an analytical conjecture regarding their structure.
\end{abstract} 

\maketitle




\section{Introduction}
The celebrated Hubbard model was introduced in 1963 as an example that a metal can become insulating by interactions \cite{G1963,H1963,K1963}. 
Anderson proposed in a highly influential paper that the two-dimensional (2D) Hubbard model is relevant for the description of high-temperature superconductivity in the cuprates \cite{A1987}.
A related model of recent interest is the so-called bilayer Hubbard model, which is a generalization of the 2D Hubbard model in which two parallel 2D lattices are stacked on top of each other and connected via nearest-neighbor hopping between the planes \cite{HVH1994}.
The bilayer model can be viewed as a step towards understanding the three-dimensional (3D)
coupling between the superconducting layers in a copper oxide high-temperature superconductor \cite{I2020}. Moreover, this model was recently realized and studied in ultracold atom experiments \cite{G2021,B2022}. 

In this paper, we investigate the phase diagram and various properties of the half-filled bilayer Hubbard model within the Hartree--Fock approximation. In particular, we obtain the weak-coupling behavior of the N\'eel temperature, as well as asymptotic formulas for the antiferromagnetic gap ratio. We also present Hartree--Fock phase diagrams for this model at zero temperature away from half-filling obtained by numerical methods, and we conjecture an analytic result about these numerical phase diagrams. 
Our results show that the 2D Hubbard model is a singular limit: the bilayer model depends on a parameter $t_\perp$ such that the limiting case $t_\perp=0$ is the 2D Hubbard model, and  some physics properties of the model are \emph{not} continuous in the limit $t_\perp\downarrow 0$. 

Our results are based on important work due to Bach, Lieb and Solovej \cite{BLS1994}, making mathematically precise folklore in the physics literature about Hartree--Fock theory for the Hubbard models on bipartite lattices, including the bilayer Hubbard model; see \cite{LLupdate} for a bridge between the physics and mathematical views on these results.

\subsection{Bilayer Hubbard model}\label{modelsec}
Let us describe the model in detail. We consider the bilayer lattice $\Lambda_L \times \{1,2\}$, where $L \geq 2$ is an even integer and $\Lambda_L$ is the two-dimensional square lattice
$$\Lambda_L = \bigg\{(x_1,x_2) \in \Z^2 \,\bigg|\, -\frac{L}{2} \leq x_1 < \frac{L}{2} \;\; \text{and} \; -\frac{L}{2} \leq x_2 < \frac{L}{2}\bigg\}.$$
We refer to $\Lambda_L \times \{1\}$ and $\Lambda_L \times \{2\}$ as the first and second layers, respectively. We use periodic boundary conditions in the $x$- and $y$-directions, i.e., we identity $\Lambda_L \times \{1,2\}$ with the union $(\Z/L\Z)^2 \times \{1,2\}$ of two torii.
We denote by $\mathcal{H} \simeq \C^{4L^2}$ the single-particle Hilbert space of $\Lambda_L \times \{1,2\}$, whose non-zero elements describe the possible states of a single electron that  can be at any of the $2L^2$ sites and can have either spin up or spin down.
The fermion Fock space $\mathcal{F} := \bigwedge \mathcal{H}$ has dimension $2^{\dim \mathcal{H}} = 16^{L^2}$.
The Hubbard model is defined by the Hamiltonian operator $H:\mathcal{F} \to \mathcal{F}$ given by $H =  H_0 + H_1$, where
\begin{align}\nonumber
& H_0 := \sum_{(\mathbf{x}, j), (\mathbf{y}, k) \in \Lambda_L \times \{1,2\}} \sum_{\sigma = \uparrow, \downarrow} t_{\mathbf{x} j \mathbf{y} k} c_{\mathbf{x} j \sigma}^\dagger c_{\mathbf{y} k \sigma}
-\mu\sum_{(\mathbf{x}, j) \in \Lambda_L \times \{1,2\}}(n_{\mathbf{x} j} -1),
	\\ \label{H0H1def}
& H_1 := U \sum_{(\mathbf{x}, j) \in \Lambda_L \times \{1,2\}} (n_{\mathbf{x} j\uparrow} -\tfrac12)(n_{\mathbf{x} j \downarrow}-\tfrac12).
\end{align}
Here the coupling parameter $U > 0$ determines the strength of the on-site repulsion, $t_{\mathbf{x} j \mathbf{y} k}$ are the hopping matrix elements, $\mu \in \R$ is the chemical potential, $c_{\mathbf{x} j \sigma}^\dagger$ is the operator on $\mathcal{F}$ which creates an electron   of spin $\sigma \in \{\uparrow, \downarrow\}$ at site $\mathbf{x} \in \Lambda_L$ in layer $j$, $c_{\mathbf{x} j \sigma}$ is the corresponding annihilation operator, $n_{\mathbf{x} j \sigma} := c_{\mathbf{x} j \sigma}^\dagger c_{\mathbf{x} j \sigma}$ is the density operator for electrons of spin $\sigma$, and $n_{\mathbf{x} j} := n_{\mathbf{x} j \uparrow}+ n_{\mathbf{x} j \downarrow}$ is the (total) density operator. 
For later reference, we also give the spin operators $\vec{s}_{\mathbf{x}j} := (s_{\mathbf{x}j1},s_{\mathbf{x}j2},s_{\mathbf{x}j3})$, 
$$
s_{\mathbf{x}j1} := c_{\mathbf{x} j \uparrow}^\dagger c_{\mathbf{x} j \downarrow}+c_{\mathbf{x} j \downarrow}^\dagger c_{\mathbf{x} j \uparrow},\quad 
s_{\mathbf{x}j2} := -ic_{\mathbf{x} j \uparrow}^\dagger c_{\mathbf{x} j \downarrow}+ic_{\mathbf{x} j \downarrow}^\dagger c_{\mathbf{x} j \uparrow},\quad 
s_{\mathbf{x}j3} := n_{\mathbf{x} j \uparrow} -n_{\mathbf{x} j \downarrow}
$$
where $i=\sqrt{-1}$.

We consider the case of nearest neighbor hopping with hopping parameter $t > 0$ within each of the two layers, and hopping parameter $t_{\perp} \geq 0$ between the two layers, i.e., for $(\mathbf{x}, j), (\mathbf{y},k) \in \Lambda_L \times \{1,2\}$ we assume that
\begin{align}\label{txybilayer}
t_{\mathbf{x} j \mathbf{y} k} 
= \begin{cases} 
-t & \text{if $\mathbf{x}$ and $\mathbf{y}$ are nearest neighbors and $j = k$},
	\\
-t_{\perp} & \text{if $\mathbf{x} = \mathbf{y}$ and $j \neq k$},
	\\
0 & \text{otherwise}.
\end{cases}
\end{align}
The limiting case $t_\perp=0$ corresponds to the 2D Hubbard model (since, in this case, the two layers are decoupled with each of them described by the 2D Hubbard model).

Since Hartree--Fock results at temperature $T$ are determined by the operator $H/T$ (see Appendix~\ref{derivationapp}), one can scale $H\to H/t$, $T\to T/t$ without changing the results: phase diagram results about the bilayer Hubbard model are determined by the parameters $U/t$, $t_\perp/t$, $\mu/t$, and $T/t$. Thus, we can set $t=1$ in the following without loss of generality.

\subsection{Hartree--Fock theory} 
We are interested in Hartree--Fock states \cite{BLS1994} of this model at fixed temperature $T\geq 0$ and fixed filling factor
\begin{equation}\label{nu} 
\nu := \frac1{2L^2}\sum_{(\mathbf{x},j)\in\Lambda_L\times\{1,2\}}\langle n_{\mathbf{x} j} -1\rangle , 
\end{equation} 
with $\langle \cdot\rangle$ the expectation value in such a state (the interested reader can find further details on Hartree--Fock theory in Appendix~\ref{derivationapp}). 
Such states are in one-to-one correspondence with functions $d(\mathbf{x},j)$ and
$\vec{m}(\mathbf{x},j)=(m_1(\mathbf{x},j),m_2(\mathbf{x},j),m_3(\mathbf{x},j))$ of $(\mathbf{x},j)\in\Lambda_L\times\{1,2\}$ 
taking values in $[-1,1]$ and the unit ball $\{\vec{m}\in\R^3; |\vec{m}|\leq 1\}$, respectively, and which correspond to 
\begin{equation}\label{HFeqs}  
d(\mathbf{x},j) = \langle n_{\mathbf{x} j} -1\rangle ,\quad \vec{m}(\mathbf{x},j) = \langle \vec{s}_{\mathbf{x} j} \rangle 
\end{equation} 
($|\vec{m}|:=\sqrt{m_1^2+m_2^2+m_3^2}$ for $\vec{m}=(m_1,m_2,m_3)\in\R^3$). 
Thus, the function $d(\mathbf{x},j)$ describes the electron density (normalized such that $d(\mathbf{x},j)=-1$, $0$, and $1$ correspond to no electron, one electron, and two electrons at site $\mathbf{x}j$, respectively), $\nu\in[-1,1]$ is the average electron density, and $\vec{m}(\mathbf{x},j)$ is the magnetization of the electrons. In particular, $\nu=0$ corresponds to half-filling: one electron per site on average. 

There are three Hartree--Fock states of particular interest called antiferromagnetic (AF) state, ferromagnetic (F) state, and paramagnetic (P) state; these states correspond to 
\begin{align*} 
\AF:&\quad d(\mathbf{x},j)=\nu,\quad\vec{m}(\mathbf{x},j) = m_{\AF} (-1)^{x_1 + x_2+j-1}  \vec{e},\\
\F:&\quad d(\mathbf{x},j)=\nu,\quad\vec{m}(\mathbf{x},j) = m_{\F}\vec{e},\\
\BI:&\quad d(\mathbf{x},j)=\nu,\quad\vec{m}(\mathbf{x},j) = \vec{0}, 
\end{align*} 
respectively, where $m_{\AF} \in (0,1]$ and $m_{\F} \in (0,1]$ are independent of $\mathbf{x} = (x_1, x_2) \in\Lambda_L$ and $j\in\{1,2\}$, and $\vec{e} \in \R^3$ is a fixed unit vector. Clearly, the P state corresponds to the limits $m_{\AF}\downarrow 0$  and $m_{\F}\downarrow 0$ of the AF and F states, respectively.  Moreover, it is known that, at half-filling, the Hartree--Fock state is either AF or P \cite{BLS1994} and, since these states are translation invariant, it is possible to derive analytic results in the thermodynamic limit $L\to\infty$ at half-filling; the mathematically results in this paper are in this category. Away from half-filling ($\nu\neq 0$), states with non-trivial functions $d(\mathbf{x},j)$ and $\vec{m}(\mathbf{x},j)$ occur, and such states are typically studied in the physics literature using large-scale numeric methods restricting to systems of finite size, $L<\infty$. However, one can detect such complicated state using Hartree--Fock theory restricted to translation invariant states as mixed states \cite{LW2007,LLupdate} and, as recently shown for the 2D Hubbard model at $T=0$ and $L\to\infty$ \cite{CLL2025}, this allows to compute phase boundaries analytically. Our numeric results are a first step towards generalizing such results to the bilayer Hubbard model.

\begin{figure}
\bigskip
\begin{center}
\begin{overpic}[width=.44\textwidth]{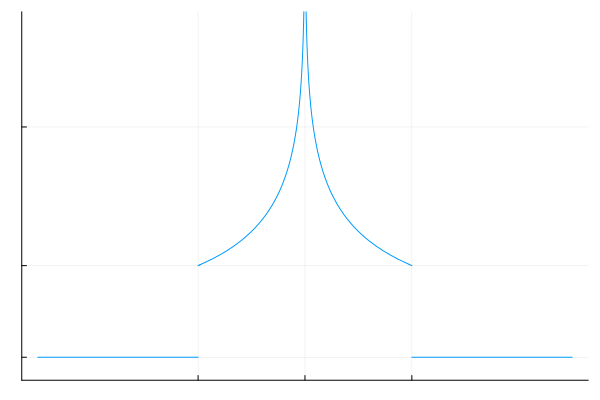}
      \put(37,69){\footnotesize $(a)$ $t_\perp = 0$}
      \put(-1,68){\footnotesize $N_{0}(\epsilon)$}
      \put(100.5,2.5){\footnotesize $\epsilon$}
       \put(29,-2){\footnotesize $^{-4}$}
      \put(50,-2){\footnotesize $^{0}$}
      \put(68,-2){\footnotesize $^{4}$}
      \put(-3.5,21.2){\footnotesize $\frac{1}{4\pi}$}
        \put(-3.5,4){\footnotesize $^{0.0}$}
       \put(-3.5,42.5){\footnotesize $^{0.2}$}
   \end{overpic}
   \hspace{0.5cm}
\begin{overpic}[width=.44\textwidth]{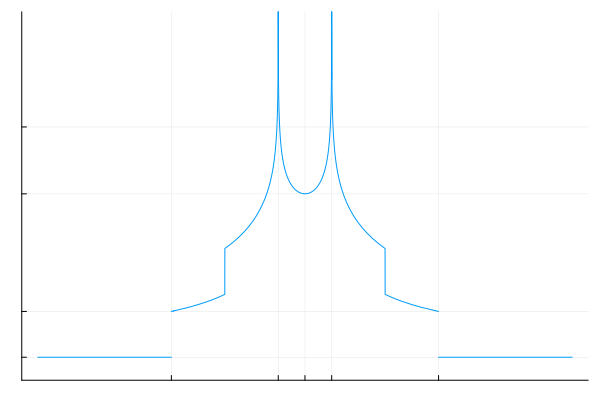}
      \put(37,69){\footnotesize $(b)$ $t_\perp = 1$}
      \put(-3,68){\footnotesize $N_{t_\perp}(\epsilon)$}
      \put(100.5,2.5){\footnotesize $\epsilon$}
      \put(22,-2){\footnotesize $^{-4-t_{\perp}}$}
      \put(42,-2){\footnotesize $^{-t_{\perp}}$}
      \put(50,-2){\footnotesize $^{0}$}
      \put(55,-2){\footnotesize $^{t_{\perp}}$}
      \put(70,-2){\footnotesize $^{4+t_{\perp}}$}
       \put(-3.5,13.4){\footnotesize $\frac{1}{8\pi}$}
       \put(-9.2,34){\tiny $^{2N_0(t_\perp)}$}
        \put(-3.5,4){\footnotesize $^{0.0}$}
       \put(-3.5,42.5){\footnotesize $^{0.2}$}
    \end{overpic}
 	\\
	\vspace{.9cm}
 \begin{overpic}[width=.44\textwidth]{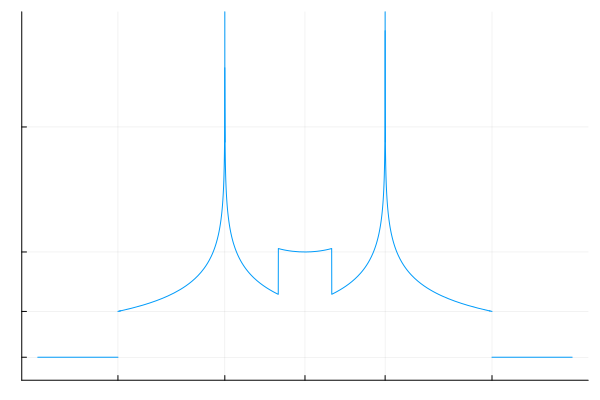}
      \put(37,69){\footnotesize $(c)$ $t_\perp = 3$}
      \put(-3,68){\footnotesize $N_{t_\perp}(\epsilon)$}
      \put(100.5,2.5){\footnotesize $\epsilon$}
      \put(13,-2){\footnotesize $^{-4-t_{\perp}}$}
      \put(35,-2){\footnotesize $^{-t_{\perp}}$}
      \put(50,-2){\footnotesize $^{0}$}
      \put(64,-2){\footnotesize $^{t_{\perp}}$}
      \put(78.5,-2){\footnotesize $^{4+t_{\perp}}$}
       \put(-3.5,13.4){\footnotesize $\frac{1}{8\pi}$}
        \put(-12,22.5){\tiny $^{2N_0(t_\perp)}$}
         \put(-3.5,4){\footnotesize $^{0.0}$}
      \put(-3.5,42.5){\footnotesize $^{0.2}$}
  \end{overpic}
   \hspace{0.5cm}
\begin{overpic}[width=.44\textwidth]{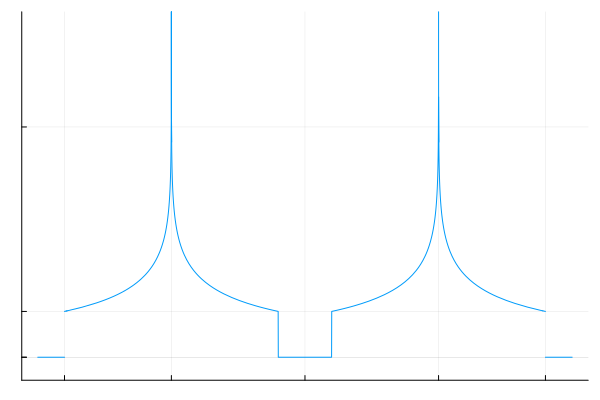}
        \put(37,69){\footnotesize $(d)$ $t_\perp = 5$}
    \put(-3,68){\footnotesize $N_{t_\perp}(\epsilon)$}
      \put(100.5,2.5){\footnotesize $\epsilon$}
      \put(5,-2){\footnotesize $^{-4-t_{\perp}}$}
      \put(25,-2){\footnotesize $^{-t_{\perp}}$}
      \put(50,-2){\footnotesize $^{0}$}
      \put(72,-2){\footnotesize $^{t_{\perp}}$}
      \put(87,-2){\footnotesize $^{4+t_{\perp}}$}
       \put(-3.5,13.4){\footnotesize $\frac{1}{8\pi}$}
        \put(-3.5,4){\footnotesize $^{0.0}$}
        \put(-3.5,42.5){\footnotesize $^{0.2}$}
   \end{overpic}
     \begin{figuretext}\label{N0Ntperpschematicfig}
       The function $N_{t_\perp}(\epsilon)$ in the case of $t = 1$ for $(a)$ $t_{\perp} = 0$, $(b)$ $t_{\perp} = 1$, $(c)$ $t_{\perp} = 3$, and $(d)$ $t_{\perp} = 5$. 
      \end{figuretext}
     \end{center}
\end{figure}

\subsection{Antiferromagnetic gap and N\'eel temperature}\label{halffillingsec}  
We recall well-known definitions and equations underlying our mathematical results at half-filling. 

Since half-filling $\nu=0$ corresponds to $\mu=0$ and only AF and P states can occur at $\mu=0$ \cite{BLS1994}, the Hartree--Fock state at half-filling can be characterized by a single parameter $m_{\AF}\in[0,1]$ that is non-zero in an AF state and zero in the P state.
We will present our results in terms of the AF order parameter $\Delta_{\AF}$ defined by
\begin{align}\label{Deltadef} 
\Delta_{\AF} := \frac{Um_{\AF}}{2},
\end{align}
which has a natural physics interpretation as the AF energy gap.
In the thermodynamic limit $L \to \infty$ of an infinitely large lattice and for fixed parameters $U>0$, $t_\perp>0$, and temperature $T>0$, $\Delta_{\AF}$ is determined by the following well-known gap equation,
\begin{align}\label{DeltaAFeq}
\frac{1}{U} =  \int_0^\infty N_{t_\perp}(\epsilon) \frac{\tanh(\frac{\sqrt{\Delta_{\AF}^2 + \epsilon^2}}{2T})}{\sqrt{\Delta_{\AF}^2 + \epsilon^2}}  
d\epsilon 
\end{align}
where $N_{t_\perp}(\epsilon)$ is the density of states for the bilayer model. The function $N_{t_\perp}(\epsilon)$ is given by (see Figure \ref{N0Ntperpschematicfig})
\begin{align}\label{bilayerdensityofstates}
N_{t_\perp}(\epsilon) = \frac{N_0(\epsilon + t_{\perp}) +N_0(\epsilon - t_{\perp})}{2},
\end{align}
where 
\begin{align}\label{2Ddensityofstates}
N_{0}(\epsilon) = \int_{[-\pi,\pi]^2} \delta\big(2 t (\cos{k_1} + \cos{k_2}) + \epsilon\big) \frac{dk_1dk_2}{(2\pi)^2}
\end{align}
is the density of states for the 2D Hubbard model. 
For parameters $U>0$, $t_\perp\geq 0$, and $T\geq 0$ such that \eqref{bilayerdensityofstates} has a solution $\Delta_{\AF} > 0$, this solution  is unique and determines $\Delta_{\AF}(U,t_\perp,T)=\Delta_{\AF}$, and we set $\Delta_{\AF}(U,t_\perp,T)=0$ for other parameters  $U>0$, $t_\perp\geq 0$ and $T\geq 0$ \cite{BLS1994}.
Derivations of the equations (\ref{DeltaAFeq})--(\ref{2Ddensityofstates}) are given in Appendix \ref{derivationapp}.

If and only if the AF phase exists, there exists a unique temperature $T_N>0$, known as the N\'eel temperature, where a phase transition from the ordered AF phase to the P phase takes place, i.e., $\Delta_{\AF}>0$ for $0\leq T<T_N$ and $\Delta_{\AF}=0$ for $T\geq T_N$. (This is  proved in Theorem~\ref{phaseth}.) 

For later reference, we also give two well-known equations implied by  (\ref{DeltaAFeq}): First,   by taking the limit as $T$ tends to $0$, one obtains an equation for the AF gap $\Delta_{\AF}=\Delta_{\AF}(U,t_\perp,0)$ at  zero temperature: 
\begin{align}\label{DeltaAFat0}
\frac{1}{U} = \int_0^\infty N_{t_\perp}(\epsilon) \frac{1}{\sqrt{\Delta_{\AF}^2 + \epsilon^2}}  d\epsilon. 
\end{align}
Second, setting $\Delta_{\AF}=0$ yields an equation determining $T_N=T_N(U,t_\perp)$: 
\begin{align}\label{TNeelequation}
\frac{1}{U} = \int_0^\infty N_{t_\perp}(\epsilon) \frac{\tanh(\frac{\epsilon}{2T_N})}{\epsilon} 
d\epsilon. 
\end{align}

\begin{figure}
\bigskip
\begin{center}
\begin{overpic}[width=.46\textwidth]{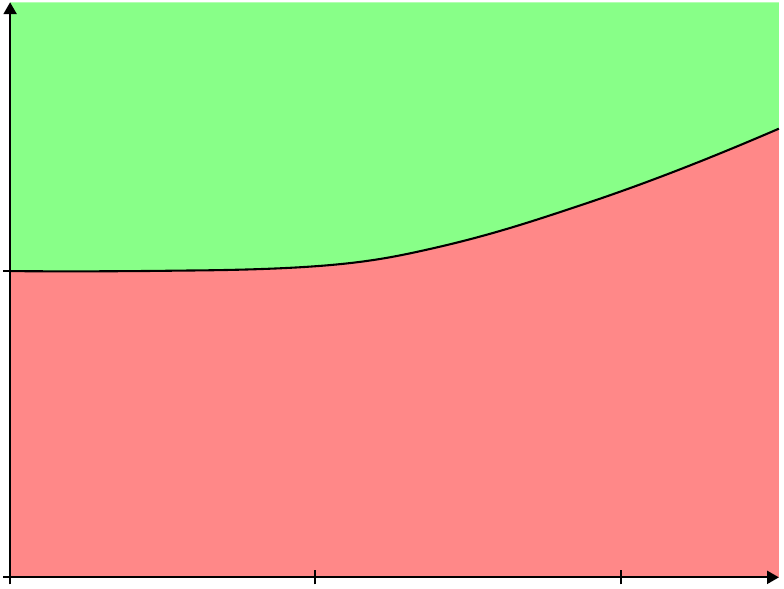}
      \put(0,77){\footnotesize $t_\perp$}
      \put(101.5,0.5){\footnotesize $U$}
       \put(-3,39.5){\footnotesize $4$}
       \put(-3,0.5){\footnotesize $0$}
       \put(0.5,-3.5){\footnotesize $0$}
       \put(39,-3.5){\footnotesize $4$}
       \put(78,-3.5){\footnotesize $8$}
       \put(32,59){\footnotesize  Paramagnetic} 
       \put(32,19){\footnotesize Antiferromagnet}
      \put(101,58){\footnotesize $U = U_c(t_\perp)$}
   \end{overpic}
      \bigskip
    \begin{figuretext}\label{phasediagramfig}
       The Hartree--Fock phase diagram for the half-filled bilayer Hubbard model at zero temperature. A paramagnetic (P) 
    phase is separated from an antiferromagnetic (AF) phase by the critical curve $U = U_c(t_\perp)$ given in (\ref{Ucdef}).
      \end{figuretext}
     \end{center}
\end{figure}

\subsection{Brief description of main results}
The following theorems are the main results of the paper.
\begin{enumerate}[$-$]

\item Theorem \ref{phaseth} establishes the Hartree--Fock phase diagram for the half-filled bilayer Hubbard model at zero temperature. Depending on the values of the parameters $U$ and $t_\perp$, the system is in a paramagnetic (P) phase or an antiferromagnetic (AF) phase, see Figure \ref{phasediagramfig}.

\item Theorem \ref{criticalth} determines the asymptotics of $\Delta_{\AF}(U, t_{\perp}, T)$ for $T = 0$ as the critical curve separating the P and AF regions is approached. 

\item Theorem \ref{3DNeelth} gives the leading asymptotic behavior of the N\'eel temperature $T_N(U,t_{\perp})$ as $U$ tends to zero for $t_{\perp} \in (0,4)$. 

\item Theorem \ref{3Dmeanfieldth} provides the asymptotics of the order parameter $\Delta_{\AF}(U, t_{\perp}, T)$ as $U \to 0$ for $t_{\perp} \in (0,4)$. Instead of stating the asymptotics for $\Delta_{\AF}$ directly, we present the asymptotics for the quotient
\begin{align}\label{hatmdef}
\hat{m}(U, t_{\perp}, T) := \frac{\Delta_{\AF}(U, t_{\perp}, T)}{T_N(U, t_{\perp})}, 
\end{align}
which we sometimes call the antiferromagnetic gap ratio. This has the advantage that the asymptotics depends on $T$ and $t_{\perp}$ only via the reduced temperature $\frac{T}{T_N(U,t_{\perp})}$.
\end{enumerate}

Moreover, in Figures \ref{roughUvsmufig} resp.\ \ref{roughUvsnufig}  we give numeric phase diagrams of the bilayer Hubbard model at zero temperature for the same fixed $t_\perp$-values as in Fig.~\ref{N0Ntperpschematicfig}: the figures show the phase as a function of $U\geq 0$ and $\mu\in\R$ resp.\ $U\geq 0$ and $\nu\in[-1,1]$ (the colored dots indicate the phase computed numerically for parameter values at the center of the dots). 
The figures suggest that the limit of the bilayer Hubbard model to the 2D Hubbard model is singular.
Our conjecture mentioned in the abstract is as follows: the transition from the P to the F phase as $\nu\to \pm 1$ occurs at $U=8\pi$ for $t_\perp>0$, jumping to $U=4\pi$ at $t_\perp=0$.

\subsection{Universality}\label{universalitysec} 
Our expansion for $\hat{m}$ in Theorem \ref{3Dmeanfieldth} is expressed in terms of the function $f_{\BCS}(y)$, which is defined for $y \in [0,1]$ as the unique solution of the equation $J(f_{\BCS}(y), y) =0$, where
\begin{align}\label{Jdef} 
J(x,y) := \int_0^\infty \bigg(\frac{\tanh(\frac{\sqrt{x^2 + \epsilon^2}}{2y})}{\sqrt{x^2 +  \epsilon^2 }} 
- \frac{\tanh(\frac{\epsilon}{2})}{\epsilon} \bigg) d\epsilon,
\end{align}
and $\tanh(\cdot/y) = 1$ for $y = 0$ by convention.
As $y$ increases from $0$ to $1$, $f_{\BCS}(y)$ decreases from $f_{\BCS}(0) = \pi e^{-\gamma} \approx 1.764$ to $f_{\BCS}(1) = 0$, where $\gamma \approx 0.5772$ is Euler's gamma constant; see Appendix~A in Ref.~\cite{LT2023} for further details.  
In Bardeen--Cooper--Schrieffer (BCS) theory, $f_{\BCS}(y)$ is important because the ratio of the superconducting gap to the critical temperature satisfies
\begin{align}\label{fracDeltaTcfBCS}
\frac{\Delta(T)}{T_c} \approx f_\BCS(T/T_c)
\end{align}
with an error that is exponentially small in the BCS coupling parameter, see e.g. \cite{L2006,LT2023}. 
The independence of the right-hand side of (\ref{fracDeltaTcfBCS}) on model parameters is a reflection of the universality of BCS theory. 
Similarly, Theorem \ref{3Dmeanfieldth} states that the function $\hat{m}$ in (\ref{hatmdef}) satisfies
\begin{align}\label{hatmfBCS}
\hat{m}(U, t_{\perp}, T) \approx f_{\BCS}(T/T_N)
\end{align}
with an error that is exponentially small in the coupling parameter $U$ of the Hubbard model.
The independence of the right-hand side of (\ref{hatmfBCS}) on model parameters reflects the universality of the Hubbard model. Note, however, that equation (\ref{hatmfBCS}) {\it only} holds when $t_{\perp} > 0$. For the two-dimensional Hubbard model (i.e., for $t_{\perp} = 0$), the universal prediction (\ref{hatmfBCS}) does not hold, but is replaced by an expansion of the form
$$\hat{m}(U, 0, T) = f_{\BCS}(T/T_N) + c_1(T/T_N) \sqrt{U} + O(U) \qquad \text{as $U \downarrow 0$}$$
for a certain function $c_1(y)$ of $y\in[0,1]$ \cite{LL2025}. 
The presence of the $U$-dependent term $c_1(T/T_N) \sqrt{U}$ on the right-hand side reflects a loss of universality in the limit $t_{\perp} \to 0$ in which the bilayer model reduces to the two-dimensional model.  
This loss of universality is another example that the limit of the bilayer Hubbard model to the 2D Hubbard model is singular. 

\subsection{Organization of the paper}
We state the main theorems in Section \ref{mainsec}. In Section \ref{phasesec}, we derive the phase diagram for the bilayer Hubbard model within the Hartree--Fock approximation. The phase diagram consists of two regions separated by a critical curve; this curve is studied in Section \ref{Ucsec}.
In Sections \ref{Neelsec} and \ref{hatmsec}, we study asymptotic properties of the N\'eel temperature $T_N$ and of the gap ratio $\hat{m}$, respectively. 
Our numeric phase diagrams are presented and discussed in Section~\ref{numericalsec}.
In Appendix \ref{derivationapp}, we include a short review of Hartree--Fock theory, including a derivation of the antiferromagnetic mean-field equation (\ref{DeltaAFeq}) for the bilayer model and the equations determining our numeric phase diagrams.

\subsection{Notation}
We use the symbols $C >0$ and $c > 0$ to denote generic constants.
The limit as $x$ tends to $a \in \R$ from above (resp. below) is written $x \downarrow a$ (resp. $x \uparrow a$).

\section{Main results}\label{mainsec}

\subsection{Hartree--Fock phase diagram}
Our first result shows that the phase diagram for the bilayer Hubbard model in the Hartree--Fock approximation consists of two regions: a paramagnetic region and an antiferromagnetic region, see Figure \ref{phasediagramfig}. 
These regions are separated by a critical curve which starts at $(U,t_\perp) = (0,4)$ and approaches the line $U \sim 2t_\perp$ for large values of $t_\perp$. More precisely, the critical curve is the graph of the function $U_c:[4,+\infty) \to [0,+\infty)$ defined by (see Figure \ref{phasediagramfig})
\begin{align}\label{Ucdef}
U_c(t_\perp) = 
\begin{cases} 
\frac{2}{\int_0^\infty N_0(\epsilon - t_{\perp}) \frac{d\epsilon}{\epsilon}}& \text{if $t_\perp \in (4,+\infty)$}.
	\\
0 & \text{if $t_\perp = 4$}.
\end{cases}
\end{align}
We let $D_{\BI}$ and $D_{\AF}$ be the paramagnetic and antiferromagnetic regions, respectively:
\begin{align}\nonumber
D_{\BI} = &\; \{(U,t_\perp) \in (0,+\infty) \times [0,+\infty) \,|\, \text{$U \leq U_c(t_\perp)$ and $4 \leq t_\perp < +\infty$}\},
	\\ \label{DBIDAFdef}
D_{\AF} = &\; \big((0,+\infty) \times [0,+\infty)\big)  \setminus D_{\BI}.
\end{align}
The following theorem summarizes the properties of the Hartree--Fock phase diagram. 

\begin{theorem}[Hartree--Fock phase diagram at half-filling]\label{phaseth}
$\;$

\begin{enumerate}[$(a)$]
\item If $(U,t_\perp) \in D_{\BI}$, then for any temperature,
equation (\ref{DeltaAFeq}) for $\Delta_{\AF}$ has no solution, i.e., $\Delta_{\AF}(U,t_\perp,T) = 0$ for all $T\geq 0$.

\item Suppose $(U,t_\perp) \in D_{\AF}$. Then equation (\ref{TNeelequation}) for the N\'eel temperature has a unique strictly positive solution $T_N = T_N(U, t_{\perp}) > 0$, and equation (\ref{DeltaAFeq}) determines a monotonically decreasing function $\Delta_{\AF}(U,t_{\perp},T)$ of $T\geq 0$ satisfying  $\Delta_{\AF}(U,t_{\perp},T)>0$ for $0\leq T<T_N(U,t_\perp)$ and $\Delta_{\AF}(U,t_{\perp},T)=0$ for $T\geq T_N(U,t_\perp)$. 
\end{enumerate}
\end{theorem}
\begin{proof}
See Section \ref{phasesec}.
\end{proof}

We show in Lemma \ref{Uclemma}  that $U_c(t_\perp)$ is a continuous function of $t_\perp \in [4,+\infty)$ such that
\begin{align}\label{Uc} 
U_c(t_\perp) = & -\frac{8\pi}{\ln(t_\perp -4)}  
 -\frac{8 \pi c_0}{\ln^2(t_\perp -4)}
+ O\bigg(\frac{1}{\ln^3(t_\perp -4)}\bigg)
\qquad \text{as $t_\perp \downarrow 4$},
\end{align}
where $c_0  \approx 3.4657$, with the exact value of $c_0$ given in (\ref{c0def}),
and such that $U_c(t_\perp) \sim 2 t_{\perp}$ for large $t_\perp$. More generally, Lemma \ref{Uclemma} provides the asymptotic expansions of $U_c(t_\perp)$ as $t_\perp \to +\infty$ and $t_\perp \downarrow 4$ to any finite order.

\subsection{Behavior near the critical curve at zero temperature}
Given $U > 0$, let $t_{\perp,c}(U)$ be the critical value of $t_\perp$ such that $(U,t_\perp) \in D_{\BI}$ if $t_\perp \geq t_{\perp,c}(U)$ and $(U,t_\perp) \in D_{\AF}$ if $t_\perp < t_{\perp,c}(U)$.
Then $t_{\perp,c}:[0,+\infty) \to [4, +\infty)$ is the inverse of $U_c$. We note in passing that \eqref{Uc} is equivalent to
\begin{align*} 
t_{\perp,c}(U) = 4 + e^{-\frac{8\pi}{U} + c_0 + O(U)} \qquad \text{as $U \downarrow 0$}.
\end{align*} 
For $t_\perp \geq t_{\perp,c}(U)$, we have $\Delta_{\AF}(U, t_{\perp}, 0) = 0$, while $\Delta_{\AF}(U, t_{\perp}, 0) > 0$ for $t_\perp < t_{\perp,c}(U)$. 
 Our next result gives the asymptotics of $\Delta_{\AF}(U, t_{\perp}, 0)$ as $t_\perp \uparrow t_{\perp,c}(U)$, i.e., as the critical curve is approached. In particular, it shows that $\Delta_{\AF}(U, t_{\perp}, 0)$ vanishes like the square root $\sqrt{t_{\perp,c}(U) - t_\perp}$ at the critical curve.

\begin{figure}
\bigskip
\begin{center}
\begin{overpic}[width=.46\textwidth]{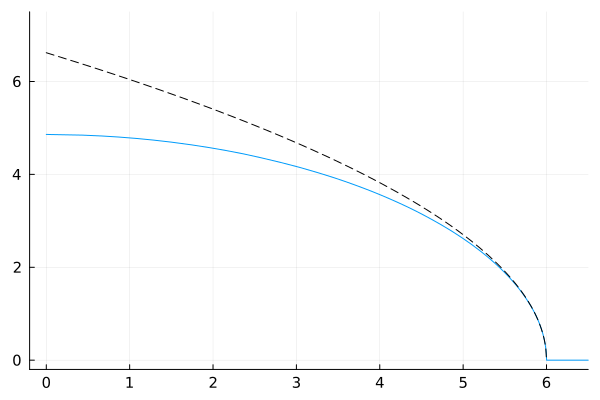}
      \put(-1,68){\footnotesize $\Delta_{\AF}(U, t_{\perp}, 0)$}
      \put(100,3.7){\footnotesize $t_\perp$}
    \end{overpic}
   \hspace{0.5cm}
\begin{overpic}[width=.46\textwidth]{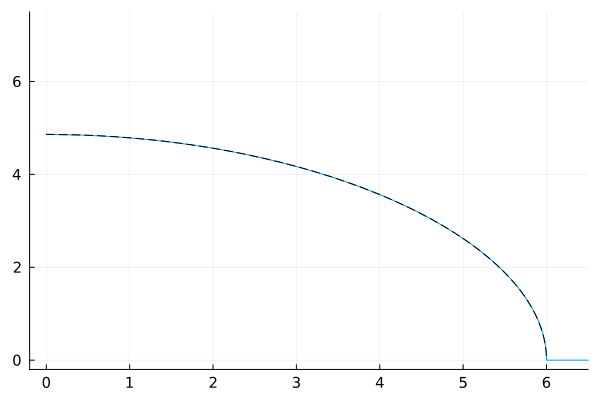}
      \put(-1,68){\footnotesize $\Delta_{\AF}(U, t_{\perp}, 0)$}
      \put(100,3.7){\footnotesize $t_\perp$}
    \end{overpic}
     \begin{figuretext}[Illustration of Theorem \ref{criticalth}]\label{criticalfig}
       The function $t_\perp \mapsto \Delta_{\AF}(U, t_{\perp}, 0)$ (solid blue) and the approximation $\sqrt{\sum_{j=1}^N d_j(U) (t_{\perp,c}(U) - t_\perp)^j}$ (dashed black) of Theorem \ref{criticalth} for $N=1$ (left) and $N= 5$ (right) in the case of $U = U_c(6) \approx 10.416$ and $t_{\perp,c}(U) = 6$.
\end{figuretext}
     \end{center}
\end{figure}

\begin{theorem}[Behavior of antiferromagnetic gap near the critical curve]\label{criticalth}
There are smooth functions $d_j:(0,+\infty) \to \R$, $j = 1, 2, \dots$, such that for any integer $N \geq 1$, the antiferromagnetic gap $\Delta_{\AF}(U, t_{\perp}, 0)$ satisfies (see Figure \ref{criticalfig})
\begin{align}\label{Deltanearcritical}
\Delta_{\AF}(U, t_{\perp}, 0)
= \sqrt{\sum_{j=1}^N d_j(U) (t_{\perp,c}(U) - t_\perp)^j + O((t_{\perp,c}(U) - t_\perp)^{N+1})}
 \qquad \text{as $t_\perp \uparrow t_{\perp,c}(U)$}
\end{align}
uniformly for $U$ in compact subsets of $(0, +\infty)$. 
The first few coefficients are given by
\begin{align}\nonumber
& d_1 =
 \frac{\int_{-4}^4 \frac{ N_0(\epsilon)}{(\epsilon + t_{\perp,c})^2}  d\epsilon}{\int_{-4}^4 \frac{ N_0(\epsilon)}{2(\epsilon + t_{\perp,c})^3}  d\epsilon} > 0, \qquad
d_2 = \frac{\int_{-4}^4 N_0(\epsilon) \big(\frac{3d_1^2}{8x^5}  - \frac{3d_1}{2x^4} + \frac{1}{x^3} \big) \big|_{x = \epsilon + t_{\perp,c}} d\epsilon}{\int_{-4}^4  \frac{N_0(\epsilon)}{2(\epsilon + t_{\perp,c})^3}  d\epsilon},
	\\\nonumber
& d_3 = - \frac{\int_{-4}^4 N_0(\epsilon) \big(\frac{5d_1^3}{16 x^7}  - \frac{15 d_1^2}{8 x^6} 
+ \frac{3 d_1(4-d_2)}{4 x^5}  
+\frac{3 d_2 - 2}{2 x^4} \big) \big|_{x = \epsilon + t_{\perp,c}} d\epsilon}{\int_{-4}^4  \frac{N_0(\epsilon)}{2(\epsilon + t_{\perp,c})^3}  d\epsilon},
	\\\nonumber
& d_4 = -\frac{1}{\int_{-4}^4  \frac{N_0(\epsilon)}{2(\epsilon + t_{\perp,c})^3}  d\epsilon} \int_{-4}^4 N_0(\epsilon) \bigg(
\frac{35 d_1^4}{128 x^9}  
- \frac{35 d_1^3}{16 x^8}  
+ \frac{15 d_1^2 (6 - d_2)}{16 x^7}  
\\\nonumber 
&  \hspace{4.5cm}  + \frac{5 d_1(3 d_2 -4)}{4 x^6}  
+ \frac{8 - 24 d_2 + 3 d_2^2 + 6 d_1 d_3}{8 x^5}  
- \frac{3 d_3}{2 x^4}  \bigg) \bigg|_{x = \epsilon + t_{\perp,c}} d\epsilon,
	\\\nonumber
& d_5 = -\frac{1}{\int_{-4}^4  \frac{N_0(\epsilon)}{2(\epsilon + t_{\perp,c})^3}  d\epsilon}
\int_{-4}^4 N_0(\epsilon) \bigg(
\frac{63 d_1^5}{256 x^{11}}-\frac{315 d_1^4}{128 x^{10}}+\frac{35 d_1^3 (8-d_2)}{32 x^9}+\frac{105 d_1^2 (d_2-2)}{16 x^8}
	\\\nonumber
& \hspace{4.5cm} +\frac{15 d_1 (d_1 d_3+d_2^2-12 d_2+8)}{16 x^7}-\frac{30 d_1 d_3+15 d_2^2-40 d_2+8}{8 x^6}
	\\ \label{djexplicit}
& \hspace{4.5cm}
+\frac{3 ((4-d_2) d_3-d_1 d_4)}{4 x^5}
 +\frac{3 d_4}{2 x^4}
     \bigg) \bigg|_{x = \epsilon + t_{\perp,c}} d\epsilon,
\end{align}
where the $U$-dependence of $d_j = d_j(U)$ and $t_{\perp,c} = t_{\perp,c}(U)$ has been suppressed for brevity.

\end{theorem}
\begin{proof}
See Section \ref{criticalsec}.
\end{proof}

\subsection{The N\'eel temperature}
Recall that $T_N(U,t_{\perp})$ reduces to the N\'eel temperature for the 2D Hubbard model when $t_{\perp} = 0$.

\begin{figure}
\bigskip
\begin{center}
\begin{overpic}[width=.46\textwidth]{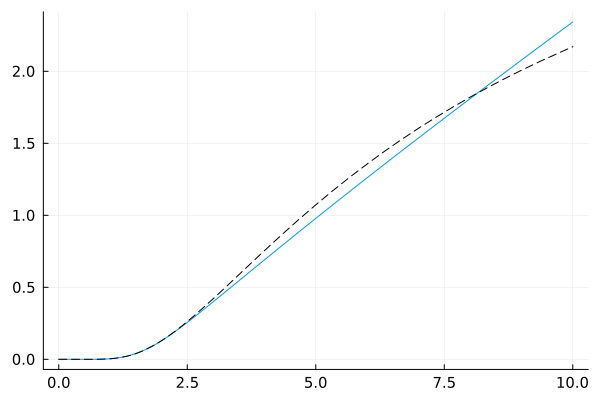}
      \put(0,68){\footnotesize $T_N(U,t_{\perp})$}
      \put(101,3.7){\footnotesize $U$}
    \end{overpic}
   \hspace{0.5cm}
\begin{overpic}[width=.46\textwidth]{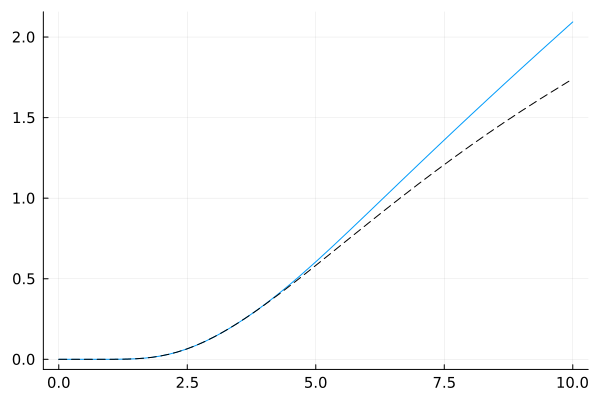}
      \put(0,68){\footnotesize $T_N(U,t_{\perp})$}
      \put(101,3.7){\footnotesize $U$}
    \end{overpic}
     \begin{figuretext}[Illustration of Theorem \ref{3DNeelth}]\label{TNfig}
       The N\'eel temperature $T_N(U,t_{\perp})$ (solid blue) and the approximation $\frac{\sqrt{4t_{\perp}(4-t_{\perp})}}{\pi e^{-\gamma}}\exp(-N_0(t_{\perp})^{-1} (1/U - b_0(t_{\perp})))$ (dashed black) of Theorem \ref{3DNeelth} for $t_{\perp} = 1$ (left) and $t_{\perp} = 3$ (right). 
\end{figuretext}
     \end{center}
\end{figure}

Our next theorem provides an asymptotic formula for the N\'eel temperature for the bilayer Hubbard model with $t_\perp \in (0, 4)$ in the small coupling limit $U \to 0$, see Figure \ref{TNfig}. 

\begin{theorem}[Small $U$ behavior of the bilayer N\'eel temperature]\label{3DNeelth}
The N\'eel temperature $T_N(U, t_{\perp})$ satisfies
\begin{align}\label{TNexpansionbilayer}
T_N(U, t_{\perp}) = \frac{\sqrt{4t_{\perp}(4-t_{\perp})}}{\pi e^{-\gamma}}e^{- \frac{1}{N_0(t_{\perp})}\big(\frac{1}{U} - b_0(t_{\perp})\big)}
\Big(1 + O(e^{-\frac{1}{N_0(t_{\perp}) U}})\Big)
 \qquad \text{as $U \downarrow 0$}
\end{align}
uniformly for $t_{\perp}$ in compact subsets of $(0, 4)$, where $\gamma$ is Euler's gamma constant and
\begin{align}\label{b0tzdef}
& b_0(t_{\perp}) := \frac{1}{2}\int_0^4 N_0(\epsilon)  \frac{d\epsilon}{|\epsilon + t_{\perp}|} + \frac{1}{2} \int_0^4 \big(N_0(\epsilon) - N_0(t_{\perp}) \big)   \frac{d\epsilon}{|\epsilon - t_{\perp}|} .
\end{align}
\end{theorem}
\begin{proof}
See Section \ref{Neelsec}.
\end{proof}

The formula of Theorem \ref{3DNeelth} should be compared with the analogous formula for the 2D Hubbard model. The bilayer Hubbard model reduces to the 2D Hubbard model in the limit $t_{\perp} \to 0$. But the limit $t_{\perp} \to 0$ is nontrivial and in fact the small $U$ behavior of the Néel temperature for the 2D Hubbard model, $T_N(U, 0)$, is qualitatively different. Indeed, it was shown in \cite[Theorem 2.4]{LL2025} that $T_N(U, 0)$ satisfies
\begin{align}\label{TNexpansion2D}
T_N(U,0)
= \frac{32}{\pi e^{-\gamma}}
e^{-\sqrt{ \frac{4 \pi^2}{U}  + a_1}}\Big(1 + O\Big(e^{-\frac{4\pi}{\sqrt{U}}}\Big)\Big) \qquad \text{as $U \downarrow 0$},
\end{align}
where
\begin{align} \label{a1def}
a_1 := & -4 \pi ^2 a_0 - \int_{0}^\infty \frac{(\ln x)^2}{\cosh^2{x}} dx + (2\ln2)^2 +(\gamma+2\ln2-\ln\pi)^2 \approx 0.3260
\end{align} 
with (see \cite{LL2026} for the evaluation of the integral)
$$a_0 := \int_{0}^{4} \bigg(N_0(\epsilon) - \frac{\ln(\frac{16}{\epsilon})}{2 \pi^2}\bigg)
\frac{1}{\epsilon}  d\epsilon = \bigg(\frac{\ln 2}{\pi}\bigg)^2 - \frac{1}{24}.$$
So while $T_N(U,t_{\perp})$ behaves like $e^{-c_1/U}$ for small $U$ whenever $t_{\perp} \in (0, 4)$, $T_N(U,0)$ behaves like $e^{-c_2/\sqrt{U}}$, where $c_1, c_2 > 0$ are known constants.

\subsection{The antiferromagnetic gap ratio}
\begin{figure}
\bigskip
\begin{center}
\begin{overpic}[width=.46\textwidth]{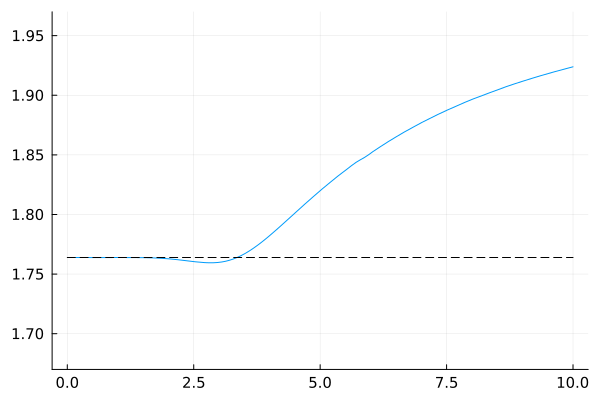}
      \put(-2,68){\footnotesize $\hat{m}(U, t_{\perp}, 0)$}
      \put(101,3.7){\footnotesize $U$}
    \end{overpic}
   \hspace{0.5cm}
\begin{overpic}[width=.46\textwidth]{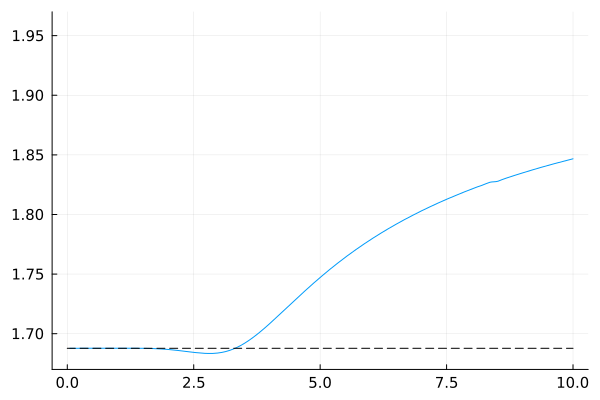}
      \put(-2,68){\footnotesize $\hat{m}(U, t_{\perp}, T_N/2)$}
      \put(101,3.7){\footnotesize $U$}
    \end{overpic}
     \begin{figuretext}[Illustration of Theorem \ref{3Dmeanfieldth}]\label{mhatfig}
       The function $U \mapsto \hat{m}(U, t_{\perp}, yT_N(U,t_{\perp}))$ (solid blue) and the approximation $f_{\BCS}(y)$ (dashed black) of Theorem \ref{3Dmeanfieldth} for $y = 0$ (left) and $y = 1/2$ (right) in the case of $t_{\perp} = 2$. 
\end{figuretext}
     \end{center}
\end{figure}

Let $\hat{m}(U, t_{\perp}, T)$ be the function defined in (\ref{hatmdef}).
Our next theorem gives the small $U$ behavior of $\hat{m}$ in the case of $t_{\perp} \in (0, 4)$, see Figure \ref{mhatfig}. 
We recall that $f_{\BCS}(y)$ is a special function of $0\leq y\leq 1$ defined in Section~\ref{universalitysec}.

\begin{theorem}[Small $U$ behavior of antiferromagnetic gap ratio]\label{3Dmeanfieldth}
The function $\hat{m}(U, t_{\perp}, T)$ satisfies
\begin{align*}
\hat{m}(U, t_{\perp}, T) = f_{\BCS}\bigg(\frac{T}{T_N(U, t_{\perp})}\bigg) + O\Big(U^{-1} e^{- \frac{2}{N_{t_\perp}(0) U}} \Big) \qquad \text{as $U \downarrow 0$}
\end{align*}
uniformly for $t_{\perp}$ in compact subsets of $(0, 4)$ and for $T \geq 0$ such that $\frac{T}{T_N(U, t_{\perp})}$ remains in a compact subset of $[0,1)$.
\end{theorem}
\begin{proof}
See Section \ref{3Dmeanfieldsubsec}. 
\end{proof} 
 
\section{Phase diagram: proof of Theorem \ref{phaseth}}\label{phasesec}

Recall that the N\'eel temperature (if it exists) is defined by equation (\ref{TNeelequation}). The next lemma shows that (\ref{TNeelequation}) has a unique strictly positive solution $T_N(U,t_\perp)$ whenever $(U,t_\perp) \in D_{\AF}$, whereas no such solution exists if $(U,t_\perp) \in D_{\BI}$, where the sets $D_{\AF}$ and $D_{\BI}$  are defined in (\ref{DBIDAFdef}) and displayed in Figure \ref{phasediagramfig}.

\begin{lemma}\label{TNlemma}
\begin{enumerate}[$(a)$]
\item \label{TNlemmaitema}
For any $t_{\perp} \in [0,4]$ and $U > 0$, there is a unique $T_N = T_N(U, t_{\perp}) > 0$ satisfying (\ref{TNeelequation}). 

\item \label{TNlemmaitemb}
For any $t_{\perp} \in (4,+\infty)$ and $U > U_c(t_\perp)$, there is a unique $T_N = T_N(U, t_{\perp}) > 0$ satisfying (\ref{TNeelequation}). If $t_{\perp} \in (4,+\infty)$ and $0 < U \leq U_c(t_\perp)$, then (\ref{TNeelequation}) has no solution in $(0,+\infty)$.

\item \label{TNlemmaitemc}
$T_N(U,t_{\perp}) \downarrow 0$ as $U \downarrow 0$ uniformly for $t_{\perp} \in [0,4]$.

\item \label{TNlemmaitemd}
$T_N(U,t_{\perp}) \downarrow 0$ as $U \downarrow U_c(t_\perp)$ uniformly for $t_{\perp}$ in compact subsets of $(4,+\infty)$.
 \end{enumerate}
\end{lemma}
\begin{proof}
For any $t_{\perp} \geq 0$, 
\begin{align}\label{ftzTdef}
f_{t_{\perp}}(T) := 2\int_0^\infty N_{t_\perp}(\epsilon) \frac{\tanh(\frac{\epsilon}{2T})}{\epsilon}  d\epsilon
\end{align}
is a continuous function of $T \in (0, +\infty)$ that tends to $0$ as $T \to +\infty$. Moreover, it is strictly decreasing because
\begin{align}\label{ftzprimeTdef}
f_{t_{\perp}}'(T) = -\int_0^\infty N_{t_\perp}(\epsilon) 
\frac{1}{T^2 \cosh^2(\frac{\epsilon}{2T})}  d\epsilon < 0  \quad \text{for all $T > 0$.}
\end{align}

$(\ref{TNlemmaitema})$ Suppose that $t_{\perp} \in [0,4]$. Then $N_{t_\perp}(\epsilon) > 0$ for all small enough $\epsilon > 0$, so $f_{t_{\perp}}(T)$ tends to $+\infty$ as $T \to 0$.
Hence the equation $f_{t_{\perp}}(T_N) = 2/U$, which is equivalent to (\ref{TNeelequation}), has a unique solution $T_N$ in $(0, +\infty)$ for any $U> 0$. 

$(\ref{TNlemmaitemb})$ Suppose that $t_{\perp} > 4$. Since $N_0(\epsilon) = 0$ for $|\epsilon|>4$, we have $\int_0^\infty N_0(\epsilon + t_{\perp}) \frac{\tanh(\frac{\epsilon}{2T})}{\epsilon}  d\epsilon= 0$, and thus
$$f_{t_{\perp}}(T) = \int_0^\infty N_0(\epsilon - t_\perp) \frac{\tanh(\frac{\epsilon}{2T})}{\epsilon}  d\epsilon.$$
Furthermore, $N_0(\epsilon - t_{\perp}) = 0$ for all small enough $\epsilon > 0$, so 
$$f_{t_{\perp}}(0) := \lim_{T \downarrow 0} f_{t_{\perp}}(T) = \int_0^\infty N_0(\epsilon - t_{\perp})  \frac{d\epsilon}{\epsilon} = \frac{2}{U_c(t_\perp)} < +\infty.$$
Hence the equation $f_{t_{\perp}}(T_N) = 2/U$, which is equivalent to (\ref{TNeelequation}), has a (necessarily unique) solution $T_N$ in $(0, +\infty)$ if and only if $\frac{2}{U} < \frac{2}{U_c(t_\perp)}$, i.e., if and only if $U > U_c(t_\perp)$.

$(\ref{TNlemmaitemc})$
Let $T_0 > 0$ and $t_\perp\geq 0$. We have from \eqref{ftzTdef}
$$
f_{t_{\perp}}(T_0) = \frac{1}{T_0}\int_0^\infty N_{t_\perp}(\epsilon) \frac{\tanh(\frac{\epsilon}{2T_0})}{\frac{\epsilon}{2T_0}}  d\epsilon \leq 
\frac{1}{T_0}\int_0^\infty N_{t_\perp}(\epsilon) d\epsilon = \frac1{2T_0}, 
$$ 
since $0\leq \frac{\tanh(x)}{x}\leq 1$ for all $x\in\R$, $N_{t_\perp}(\epsilon)\geq 0$ for all $\epsilon\geq 0$, and $\int_0^\infty N_{t_\perp}(\epsilon)d\epsilon =\frac12$. 
Hence, for each $t_{\perp} \in [0,4]$, if $\frac{2}{U} > \frac{1}{2T_0}$ (i.e., if $U < 4T_0$), then $T_N(U, t_{\perp}) < T_0$. Since $T_0 > 0$ was arbitrary, we conclude that $T_N(U,t_{\perp}) \downarrow 0$ uniformly for $t_{\perp} \in [0,4]$ as $U \downarrow 0$.

$(\ref{TNlemmaitemd})$
There is a constant $c > 0$ such that
\begin{align*}
f_{t_\perp}(0) - f_{t_\perp}(T) 
& = 2\int_0^\infty N_0(\epsilon - t_\perp) \frac{1 - \tanh(\frac{\epsilon}{2T})}{\epsilon} d\epsilon
= 4\int_{t_\perp - 4}^{t_\perp + 4}  \frac{N_0(\epsilon - t_\perp)}{1 + e^{\frac{\epsilon}{T}}} \frac{d\epsilon}{\epsilon} 
	\\
& \geq \frac{c}{(1 + e^{\frac{t_\perp + 4}{T}})(t_\perp + 4)}
\end{align*}
for all $T > 0$ and $t_\perp > 4$.
Consequently, using also that $ f_{t_\perp}(T_N(U,t_\perp)) = \frac{2}{U}$, we have
\begin{align}\label{2Uc2U}
\frac{2}{U_c(t_\perp)} - \frac{2}{U} = f_{t_\perp}(0) - f_{t_\perp}(T_N(U,t_\perp)
\geq \frac{2c}{(1 + e^{\frac{t_\perp + 4}{T_N(U,t_\perp)}})(t_\perp + 4)}
\end{align}
 for all $t_\perp > 4$ and $U > U_c(t_\perp)$.
Let $[a,b]$ be a compact subset of $(4,+\infty)$. As $U \downarrow U_c(t_\perp)$, we have $\frac{2}{U_c(t_\perp)} - \frac{2}{U} \to 0$ uniformly for $t_\perp \in [a,b]$, and hence (\ref{2Uc2U}) implies that $T_N(U,t_\perp) \downarrow 0$ uniformly for $t_\perp \in [a,b]$.
\end{proof}

The next lemma shows that there exists a unique strictly positive solution $\Delta_{\AF}$ of (\ref{DeltaAFeq}) if and only if $(U,t_\perp) \in D_{\AF}$ and $T \in [0, T_N(U,t_\perp))$. 
It also shows that it always holds that $0\leq \Delta_{\AF}\leq U/2$, i.e., that $0\leq m_{\AF}\leq 1$; the latter is expected since $m_{\AF}$ is the  magnitude of the spin expectation value.

\begin{lemma}\label{mAFlemma}
\begin{enumerate}[$(a)$]
\item \label{mAFlemmaitema}
For any $(U,t_{\perp}) \in D_{\AF}$, equation (\ref{DeltaAFeq}) has a unique strictly positive solution $\Delta_{\AF} = \Delta_{\AF}(U,t_{\perp},T) > 0$ if $T \in [0, T_N(U,t_{\perp}))$,  whereas if $T \geq T_N(U,t_{\perp})$ then $\Delta_{\AF} = \Delta_{\AF}(U,t_{\perp},T) = 0$.

\item \label{mAFlemmaitemb}
$\Delta_{\AF}(U,t_{\perp},T) \in [0,U/2)$ for all $(U,t_{\perp}) \in D_{\AF}$ and all $T \geq 0$.

\item \label{mAFlemmaitemc} The function $\Delta_{\AF} = \Delta_{\AF}(U,t_{\perp},T)$ in $\eqref{mAFlemmaitema}$ is nonincreasing in $ T \geq 0$.  
\end{enumerate}
\end{lemma}
\begin{proof}
Fix $t_{\perp} \in [0,+\infty)$. If $t_{\perp} \in [0,4]$, suppose that $U > 0$, while if $t_{\perp} \in (4,+\infty)$, suppose that $U > U_c(t_\perp)$. Define the function $F_T(\Delta)$ for $\Delta \geq 0$ by
\begin{align}\label{Fmdef}
F_T(\Delta) = 2\int_0^\infty N_{t_\perp}(\epsilon) 
\frac{\tanh(\frac{\sqrt{\Delta^2 + \epsilon^2}}{2T})}{\sqrt{\Delta^2 + \epsilon^2}} 
 d\epsilon.
\end{align}

To prove assertions $(\ref{mAFlemmaitema})$, is enough to show that the equation $F_T(\Delta) = 2/U$ has a unique solution $\Delta > 0$ whenever $T \in [0, T_N(U,t_{\perp}))$ and that no such solution exists if $T \geq T_N(U,t_{\perp})$.
The function $F_T$ is continuous $[0,+\infty) \to [0,+\infty)$ and obeys $F_T(\Delta) \to 0$ as $\Delta \to +\infty$. Moreover, $F_T(0)$ equals the right-hand side of (\ref{ftzTdef}), which is a decreasing function of $T \in [0,+\infty)$ by (\ref{ftzprimeTdef}). 
The N\'eel temperature $T_N = T_N(U,t_{\perp})$ is defined by the equation $F_{T_N}(0) = 2/U$.
We infer that if $T < T_N$, then $F_T(0) > 2/U$, and so the equation $F_T(\Delta) = 2/U$ has at least one solution $\Delta$ in $(0, +\infty)$. Since
\begin{align*}
F_T'(\Delta) = &\; 2\int_0^\infty N_{t_\perp}(\epsilon) \frac{\Delta}{(\Delta^{2} + \epsilon^2)^{3/2}} \frac{x - \sinh(x)\cosh(x)}{\cosh^2(x)}\bigg|_{x = \frac{\sqrt{\Delta^{2} + \epsilon^2}}{2T}} d\epsilon
\end{align*}
and $x - \sinh(x)\cosh(x) < 0$ for all $x > 0$, we have $F_T'(\Delta) < 0$ for $\Delta \in (0, +\infty)$ so the solution is unique. This proves $(\ref{mAFlemmaitema})$.

If $\Delta \geq U/2$, then
$$F_T(\Delta) < \frac{2}{\Delta}  \int_0^\infty N_{t_\perp}(\epsilon)  d\epsilon
\leq \frac{2}{U} \int_\R N_{t_\perp}(\epsilon) d\epsilon
= \frac{2}{U},$$
so no $\Delta \geq U/2$ can fulfill the equation $F_T(\Delta) = 2/U$. This proves $(\ref{mAFlemmaitemb})$.

To prove $(\ref{mAFlemmaitemc})$, we show that  $\frac{\partial \Delta_{\AF}}{\partial T}\leq 0$ for all $T\geq 0$. This is obvious for $T\geq T_N(U,t_\perp)$. For $0\leq T< T_N(U,t_\perp)$, this follows from $\frac{\partial \Delta_{\AF}}{\partial T} = -\frac{\partial F_T(\Delta_{\AF})}{\partial T}/F_{T}'(\Delta_{\AF})\leq 0$ obtained  by taking the total $T$-derivative of $\frac1U = F_T(\Delta_{\AF})$, together with $F'_T(\Delta_{\AF})<0$ and $\frac{\partial F_T(\Delta)}{\partial T}< 0$ for $\Delta \in (0, +\infty)$; the former was already shown above, and the latter follows from the relation
$$
\frac{\partial F_T(\Delta)}{\partial T} = -\frac1{T^2}\int_0^\infty N_{t_\perp}(\epsilon) 
\frac{1}{\cosh^2(\frac{\sqrt{\Delta^2 + \epsilon^2}}{2T})}  d\epsilon < 0
$$
implied by \eqref{Fmdef}. 
\end{proof}

Lemma \ref{mAFlemma} implies that $\Delta_{\AF} = 0$ is the only solution of the Hartree--Fock equations if $T \geq T_N$. On the other hand, for any $T \in [0, T_N)$, there are two distinct solutions in $[0,U/2]$, namely $0$ and $\Delta_{\AF}(U,t_{\perp},T) \in (0, U/2)$. The next lemma compares the free energy densities of these solutions.
The statement of the lemma involves the thermodynamic limit of the Hartree--Fock function $\mathcal{G}$ restricted to AF states, which we now denote by $\mathcal{G}_T(\Delta)$ to make the dependence on $T$ and $\Delta$ explicit (see (\ref{mathcalGAF})):
\begin{align*}
\mathcal{G}_T(\Delta) := &\; \begin{cases}
\frac{\Delta^2}{U} - 2 T \int_0^{+\infty} N_{t_\perp}(\epsilon) \ln\big( 2 \cosh\big(\frac{\sqrt{\Delta^2 + \epsilon^2}}{2 T}\big)\big) d\epsilon, & 0 < T < \infty,
	\\
\frac{\Delta^2}{U} -\int_0^{+\infty} N_{t_\perp}(\epsilon)  \sqrt{\Delta^2 + \epsilon^2}  d\epsilon, & T = 0.
\end{cases}
\end{align*}
If $\Delta_\AF$ is a solution of the Hartree--Fock equation (\ref{DeltaAFeq}), then the free energy density of the corresponding AF state equals $\mathcal{G}_T(\Delta)$, see e.g.  \cite[Eq. (20)]{LLupdate}. The following lemma therefore shows that the state corresponding to $\Delta_{\AF}(U,t_{\perp},T) > 0$ has lower energy than the state corresponding to $0$ whenever it exists.

\begin{lemma}\label{Gminlemma}
Let $(U, t_{\perp}) \in D_{\AF}$. If $T \in [0, T_N(U,t_{\perp}))$, then $\Delta = \Delta_{\AF}(U,t_{\perp},T) > 0$ is the unique point at which $\mathcal{G}_T$ attains its minimum on $[0,U/2]$.
If $T \geq T_N(U,t_{\perp})$, then $\Delta = 0$ is the unique point at which $\mathcal{G}_T$ attains its minimum on $[0,U/2]$.
\end{lemma}
\begin{proof}
Fix $(U, t_{\perp}) \in D_{\AF}$. We have, for any $T \geq 0$,
$$\mathcal{G}_T'(\Delta) = \Delta \bigg(\frac{2}{U} - \frac{F_T(\Delta)}{2}\bigg),
$$
where $F_T(\Delta)$ is the function in (\ref{Fmdef}). 
Suppose first that $T \in [0, T_N(U,t_{\perp}))$. In this case, the proof of Lemma \ref{mAFlemma} shows that $F_T(\Delta) > 2/U$ for $\Delta \in [0, \Delta_{\AF}(U,t_{\perp},T))$ and $F_T(\Delta) < 2/U$ for $\Delta \in (\Delta_{\AF}(U,t_{\perp},T), U/2]$. Hence $\Delta_{\AF}(U,t_{\perp},T) \in (0,U/2)$ is the unique critical point of $\mathcal{G}_T$ on $(0,U/2)$ and $\mathcal{G}_T$ is decreasing (resp. increasing) to the left (resp. right) of $\Delta_{\AF}(U,t_{\perp},T)$, so $\Delta_{\AF}(U,t_{\perp},T)$ is indeed the minimizer of $\mathcal{G}_T$. 

If $T \geq T_N(U,t_{\perp})$, then we instead have $\mathcal{G}_T'(0) = 0$ and $\mathcal{G}_T'(\Delta) > 0$ for $\Delta \in (0,U/2)$, so $0$ is the unique minimizer of $\mathcal{G}_T$.
\end{proof}

\section{The critical curve}\label{Ucsec}
In this section, we study the function $U_c(t_\perp)$. In particular, we derive asymptotic expansions of $U_c(t_\perp)$ as  $t_\perp$ tends to $4$ and to $+\infty$. It turns out that to describe the behavior of $U_c(t_\perp)$ as  $t_\perp \downarrow 4$ we need the expansion of the two-dimensional density of states $N_0(\epsilon)$ as $\epsilon \uparrow 4$. The next lemma expresses $N_0(\epsilon)$ for $\epsilon \in (0,4)$ as the restriction of a power series based at $4$ with radius of convergence equal to $4$; in particular, it provides the desired asymptotic expansion of $N_0(\epsilon)$ as $\epsilon \uparrow 4$ to all orders.

\begin{figure}
\bigskip
\bigskip
\begin{center}
\begin{overpic}[width=.46\textwidth]{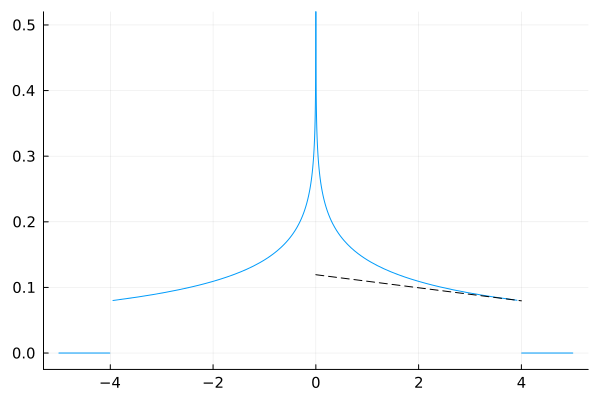}
      \put(2,68){\footnotesize $N_0(\epsilon)$}
      \put(101,3.7){\footnotesize $\epsilon$}
    \end{overpic}
   \hspace{0.5cm}
\begin{overpic}[width=.46\textwidth]{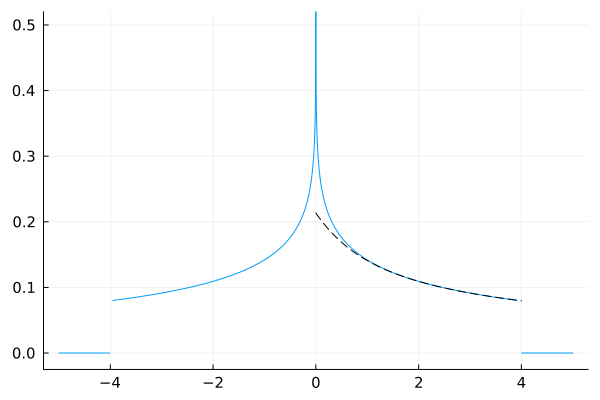}
      \put(2,68){\footnotesize $N_0(\epsilon)$}
      \put(101,3.7){\footnotesize $\epsilon$}
    \end{overpic}
     \begin{figuretext}[Illustration of Lemma \ref{N0near4lemma}]\label{N0fig}
       Left: The function $N_0(\epsilon)$ (solid blue) and the approximation $\frac{1}{4 \pi}
+\frac{4-\epsilon}{32 \pi}$ (dashed black).
       Right: The function $N_0(\epsilon)$ (solid blue) and its approximation obtained by dropping the $O((4-\epsilon)^{9})$-term in (\ref{N0expansionfirstfew}) (dashed black). 
      \end{figuretext}
     \end{center}
\end{figure}

\begin{lemma}\label{N0near4lemma}
For $\epsilon \in (0, 4)$, it holds that
\begin{align}\label{N0near4}
N_0(\epsilon) =  \sum_{j=0}^\infty a_j (4-\epsilon)^{j},
\end{align}
where
\begin{align}\label{ajdef}
a_j := \frac{(-1)^j}{4^{1+j}\pi j!} \bigg(\sum_{m=0}^j \frac{\Gamma(j-m+\frac{1}{2}) \Gamma(m+\frac{1}{2})}{(j-m)! m! \Gamma(\frac{1}{2}-(j-m)) \Gamma(\frac{1}{2}-m)} \bigg), \qquad j = 0, 1, \dots.
\end{align}
In particular, retaining the first several terms in (\ref{N0near4}), we find (see Figure \ref{N0fig})
\begin{align}\nonumber
N_0(\epsilon) = &\; \frac{1}{4 \pi}
+\frac{\big(\frac{4-\epsilon}{32}\big)}{\pi}
+\frac{5 \big(\frac{4-\epsilon}{32}\big)^2}{\pi}
+\frac{28 \big(\frac{4-\epsilon}{32}\big)^3}{\pi}
+\frac{169 \big(\frac{4-\epsilon}{32}\big)^4}{\pi}
+\frac{1076 \big(\frac{4-\epsilon}{32}\big)^5}{\pi}
+\frac{7124 \big(\frac{4-\epsilon}{32}\big)^6}{\pi}
	\\ \label{N0expansionfirstfew}
& +\frac{48560 \big(\frac{4-\epsilon}{32}\big)^7}{\pi}
+\frac{338377 \big(\frac{4-\epsilon}{32}\big)^8}{\pi}
+ O((4-\epsilon)^{9})
\qquad \text{as $\epsilon \uparrow 4$}.
\end{align}
\end{lemma}
\begin{proof}
We know from \cite[Eq. (4.8)]{LL2025} that
\begin{align}\label{N0w1w2}
N_0(\epsilon)
   = \frac{1}{\pi^2(4-\epsilon)} \int_{0}^{1} \frac{1}{\sqrt{v} \sqrt{v-1 + w_2} \sqrt{w_2 - v} \sqrt{1-v}} dv \qquad \text{for $\epsilon \in (0,4)$},
\end{align}   
where principal branches are used and $w_2 := \frac{4}{4-\epsilon}$.
The power series representation
$$\frac{1}{\sqrt{1-x}} = \sum_{n=0}^\infty \frac{\sqrt{\pi } (-1)^n x^n}{n!
   \Gamma \left(\frac{1}{2}-n\right)}$$
converges uniformly for $x$ in compact subsets of the open unit disk. Using also that
$$0<\frac{1-v}{w_2} < 1-\frac{\epsilon}{4}\quad \text{and} \quad 0<\frac{v}{w_2} < 1-\frac{\epsilon}{4} \quad \text{for $v \in (0,1)$ and $\epsilon \in (0,4)$},$$ 
we obtain, for each $\epsilon \in (0,4)$,
\begin{align*}
N_0(\epsilon)
& = \frac{1}{\pi^2(4-\epsilon)w_2} \int_{0}^{1} \frac{1}{\sqrt{v} \sqrt{1 - \frac{1-v}{w_2}} \sqrt{1 - \frac{v}{w_2}} \sqrt{1-v}} dv
	\\
& = \frac{1}{4\pi^2} \sum_{n,m=0}^\infty \frac{\sqrt{\pi } (-1)^n }{n! \Gamma(\frac{1}{2}-n)} \frac{\sqrt{\pi } (-1)^m }{m!
   \Gamma(\frac{1}{2}-m)} w_2^{-(n+m)}  \int_{0}^{1} \frac{(1-v)^n v^{m}}{\sqrt{v} \sqrt{1-v}} dv.
\end{align*}
Since
$$\int_{0}^{1} \frac{(1-v)^n v^{m}}{\sqrt{v} \sqrt{1-v}} dv = \frac{\Gamma(n+\frac{1}{2}) \Gamma(m+\frac{1}{2})}{(m+n)!},$$
this yields
\begin{align*}
N_0(\epsilon)
& = \frac{1}{4^{1+n+m}\pi}  \sum_{n,m=0}^\infty  \frac{(-1)^{n+m}}{n! m! \Gamma(\frac{1}{2}-n) \Gamma(\frac{1}{2}-m)} \frac{\Gamma(n+\frac{1}{2}) \Gamma(m+\frac{1}{2})}{(m+n)!}
   (4-\epsilon)^{n+m}
\end{align*}
for $\epsilon \in (0,4)$. Letting $j = n+m$, we find the desired representation.
\end{proof}

\begin{remark}
An alternative proof of Lemma \ref{N0near4lemma} uses that $N_0(\epsilon) = \frac{2}{\pi^2 |\epsilon|} K\big(1 - \frac{16}{\epsilon^2}\big)$ for $\epsilon \in [-4,4]$  (see e.g. \cite{LL2025}) together with known properties of the complete elliptic integral $K$.
\end{remark}

\begin{remark}
The sum over $m$ in (\ref{N0near4}) can be expressed in terms of the hypergeometric ${}_3F_2$-function, which yields
\begin{align*}
N_0(\epsilon) =  \sum_{j=0}^\infty \frac{(-1)^j \Gamma(j+\frac{1}{2})}{4^{1+j}\pi (j!)^2 \Gamma(\frac{1}{2}-j) } 
\, {}_3F_2\left(\frac{1}{2},\frac{1}{2},-j;\frac{1}{2}-j,\frac{1}{2}-j;-1\right) (4-\epsilon)^{j}
\quad \text{for $\epsilon \in (0,4)$},
\end{align*}
where 
$${}_3F_2\left(\frac{1}{2},\frac{1}{2},-j;\frac{1}{2}-j,\frac{1}{2}-j;-1\right)
= \sum_{k=0}^\infty \frac{(\frac{1}{2})_k (\frac{1}{2})_k (-j)_k}{(\frac{1}{2}-j)_k (\frac{1}{2}-j)_k} \frac{(-1)^k}{k!}$$
and $(a)_k := \Gamma(a+k)/\Gamma(a)$ denotes the Pochhammer symbol.
\end{remark}

\begin{figure}
\bigskip
\begin{center}
\begin{overpic}[width=.46\textwidth]{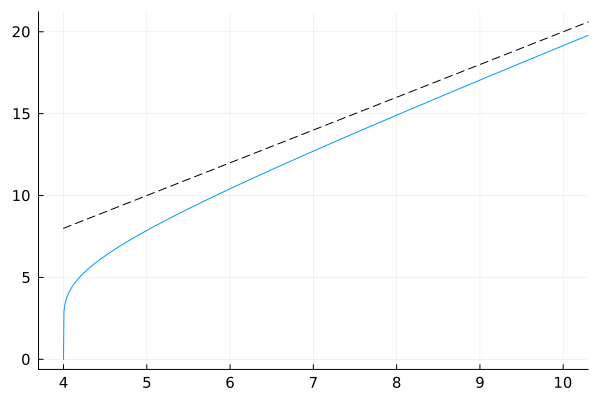}
      \put(0,68){\footnotesize $U_c(t_\perp)$}
      \put(101,3.7){\footnotesize $t_\perp$}
    \end{overpic}
   \hspace{0.5cm}
\begin{overpic}[width=.46\textwidth]{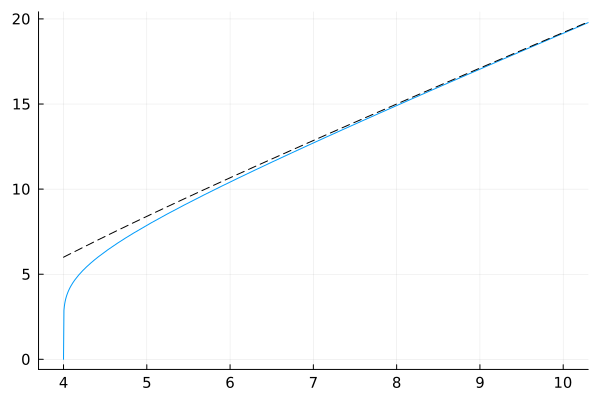}
      \put(0,68){\footnotesize $U_c(t_\perp)$}
      \put(101,3.7){\footnotesize $t_\perp$}
    \end{overpic}
     \begin{figuretext}[Illustration of part $(\ref{Ucatinftyitem})$ of Lemma \ref{Uclemma}]\label{Ucfig}
       Left: The function $U_c(t_\perp)$ (solid blue) and the leading asymptotic approximation $2 t_\perp$ (dashed black).
       Right: The function $U_c(t_\perp)$ (solid blue) and the asymptotic approximation given in  (\ref{Ucatinftyfirstfew}) (dashed black).
            \end{figuretext}
     \end{center}
\end{figure}

\begin{figure}
\bigskip
\bigskip
\begin{center}
\begin{overpic}[width=.46\textwidth]{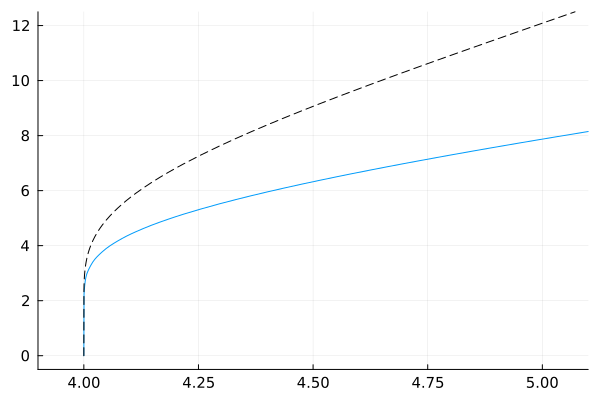}
      \put(0,68){\footnotesize $U_c(t_\perp)$}
      \put(101,3.7){\footnotesize $t_\perp$}
    \end{overpic}
   \hspace{0.5cm}
\begin{overpic}[width=.46\textwidth]{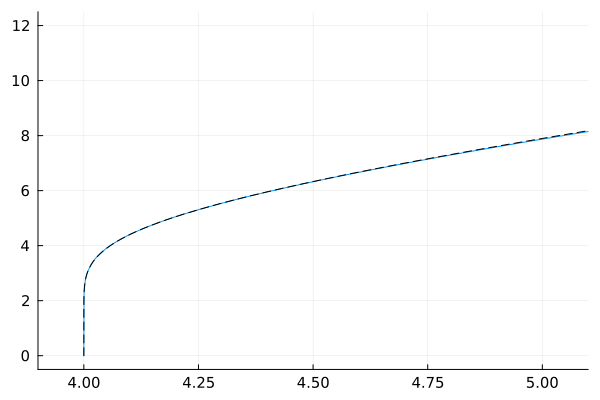}
      \put(0,68){\footnotesize $U_c(t_\perp)$}
      \put(101,3.7){\footnotesize $t_\perp$}
    \end{overpic}
     \begin{figuretext}[Illustration of part $(\ref{Ucat4item})$ of Lemma \ref{Uclemma}]\label{Ucnear4fig}
       Left: The function $U_c(t_\perp)$ (solid blue) and the leading asymptotic approximation $-8\pi/\ln(\frac{t_\perp -4}{8})$ (dashed black).
       Right: The function $U_c(t_\perp)$ (solid blue) and the asymptotic approximation given in  (\ref{Ucat4firstfew}) (dashed black).
      \end{figuretext}
     \end{center}
\end{figure}

The next lemma shows that $U_c(t_\perp)$ behaves like the linear function $2 t_{\perp}$ for large $t_\perp$, see Figure \ref{Ucfig}. It also provides the asymptotic expansion of $U_c(t_\perp)$ to any finite order as $t_\perp \downarrow 4$, see Figure \ref{Ucnear4fig}. In particular, it shows that $U_c(t_\perp)$ is a continuous function of $t_\perp \in [4,+\infty)$ with the leading behavior $-\frac{8\pi}{\ln(\frac{t_\perp -4}{8})}$ as $t_\perp \downarrow 4$.

\begin{lemma}\label{Uclemma}
The function $U_c(t_\perp)$ has the following properties:
\begin{enumerate}[$(a)$]
\item \label{Ucatinftyitem}
For any integer $N \geq 0$, it holds that
\begin{align}\label{Ucatinfty}
U_c(t_\perp) = 2 t_{\perp}\bigg(1 + \sum_{j=1}^{N} \bigg(- \sum_{k=1}^{N} \frac{2}{t_\perp^{2k}} \binom{2k}{k}^2\bigg)^j \bigg) + O\bigg(\frac{1}{t_\perp^{2N+1}}\bigg)  \qquad \text{as $t_\perp \to +\infty$}.
\end{align}
In particular, (\ref{Ucatinfty}) with $N=1$ gives (see Figure \ref{Ucfig})
\begin{align}\label{Ucatinftyfirstfew}
U_c(t_\perp) = 2 t_{\perp} 
-\frac{8}{t_\perp}
+ O\bigg(\frac{1}{t_\perp^{3}}\bigg)  \qquad \text{as $t_\perp \to +\infty$}.
\end{align}

\item \label{Ucat4item}
For any integer $N \geq 0$, it holds that
\begin{align}\label{Ucat4}
U_c(t_\perp) = \frac{2}{\sum_{j=0}^{N-1} \big(b_j(N) - a_j \ln(t_\perp - 4)\big) (4 - t_\perp)^j} + O((t_\perp - 4)^N)
\qquad \text{as $t_\perp \downarrow 4$},
\end{align}
where $\{a_j\} \subset \R$ are given by (\ref{ajdef}), and $\{b_j(N)\} \subset \R$ are defined by
\begin{align}\label{bjdef}
b_j(N) :=  a_j \ln{8} 
+ \sum_{\substack{k=0 \\ k \neq j}}^N a_k \frac{8^{k-j}}{k-j}
+ \int_0^8 \bigg(N_0(4-x) -  \sum_{l=0}^N a_l  x^l\bigg) \frac{1}{x^{j+1}} dx.
\end{align}

In particular, $U_c(t_\perp)$ is a continuous function of $t_\perp \in [4,+\infty)$, and (\ref{Ucat4}) with $N=1$ gives
\begin{align}\nonumber
U_c(t_\perp) =& -\frac{8 \pi }{\ln(\frac{t_\perp -4}{8}) -4 \pi \int_0^8 (N_0(4-x) -  a_0 - a_1x) \frac{1}{x} dx -1} + O(t_\perp -4)
	\\
= & -\frac{8\pi}{\ln(t_\perp -4)}  
 -\frac{8 \pi c_0}{\ln^2(t_\perp -4)}
+ O\bigg(\frac{1}{\ln^3(t_\perp -4)}\bigg)
\qquad \text{as $t_\perp \downarrow 4$},
	\\ \label{c0def}
c_0 := & \; 4 \pi  \int_0^8 \bigg(N_0(4-x) - \frac{1}{4\pi} - \frac{x}{32\pi}\bigg) \frac{1}{x} dx +1+\ln{8} \approx 3.4657,
\end{align}
while (\ref{Ucat4}) with $N=2$ gives the better approximation (see Figure \ref{Ucnear4fig})
\begin{align}\nonumber
U_c(t_\perp) = &\; \frac{64 \pi }{32 \pi \int_0^8 (N_0(4-x) -  \sum_{l=0}^2 a_l  x^l) \frac{1}{x} dx -8 \ln(\frac{t_\perp -4}{8})+13}
	\\\nonumber
& +\frac{16 \pi  (128 \pi \int_0^8 (N_0(4-x) -  \sum_{l=0}^2 a_l  x^l) \frac{1}{x^{2}} dx -4
   \ln(\frac{t_\perp -4}{8})+1)}{(32 \pi  \int_0^8 (N_0(4-x) -  \sum_{l=0}^2 a_l  x^l) \frac{1}{x} dx -8
   \ln(\frac{t_\perp -4}{8})+13)^2}(t_\perp -4)
   	\\ \label{Ucat4firstfew}
&  + O((t_\perp - 4)^2)
\qquad \text{as $t_\perp \downarrow 4$},
\end{align}
\end{enumerate}
\end{lemma}

\begin{proof}
We write $U_c(t_\perp) = 2/V(t_\perp)$, where $V(t_\perp) := \int_0^\infty N_0(\epsilon - t_{\perp}) \frac{d\epsilon}{\epsilon}$.

$(\ref{Ucatinftyitem})$ 
Let $N \geq 0$ be an integer.
Since $N_0(\epsilon) = 0$ for $|\epsilon|>4$, we find after changing variables to $x = \epsilon - t_{\perp}$ that
$$V(t_\perp)= \int_{t_{\perp} - 4}^{t_{\perp} + 4} N_0(\epsilon - t_{\perp}) \frac{d\epsilon}{\epsilon}
= \int_{-4}^4 N_0(x) \frac{dx}{x + t_\perp}$$
for any $t_\perp > 4$. Since the function $N_0(\epsilon)$ is even, $\int_{-4}^4 N_0(x) x^k dx = 0$ for any odd $k$.
Thus, using also that $\int_{-4}^4 N_0(x)  = 1$,
\begin{align}\nonumber
V(t_\perp) & = \int_{-4}^4 N_0(x) \frac{1}{t_\perp}\bigg(1 + \sum_{l=1}^{2N+1} (-1)^{l} \frac{x^l}{t_\perp^l} + O\bigg(\frac{x^{2N+2}}{t_\perp^{2N+2}}\bigg)\bigg) dx
	\\ \label{Vtperp}
& = \frac{1}{t_\perp}\bigg(1 + \sum_{k=1}^{N} \frac{2}{t_\perp^{2k}} \int_{0}^4 N_0(x) x^{2k} dx + O\bigg(\frac{1}{t_\perp^{2N+2}}\bigg)\bigg) 
\qquad \text{as $t_\perp \to +\infty$}.
\end{align}
By Lemma 3.1 in \cite{LL2026}, we have, using the Legendre duplication formula in the second step,
\begin{align*} 
\int_{-4}^4 N_0(x) x^{2k} dx = 2 \frac{4^{2k+1}}{8\pi}\left(\frac{\Gamma(k+\tfrac12)}{\Gamma(k+1)} \right)^2 
= \binom{2k}{k}^2. 
\end{align*}
Substituting this expression into (\ref{Vtperp}) and recalling that $U_c(t_\perp) = 2/V(t_\perp)$, assertion $(\ref{Ucatinftyitem})$ follows. 

$(\ref{Ucat4item})$ 
Let $s := t_\perp - 4$ with $t_\perp > 4$ and let $N \geq 0$ be an integer. Using the fact that $N_0(\epsilon) = 0$ for $|\epsilon|>4$, the change of variables $\epsilon = s + x$, and the evenness of $N_0$, we can write
$$V(t_\perp) = \int_{s}^{s+8} N_0(\epsilon - s - 4) \frac{d\epsilon}{\epsilon}
= \int_0^8 N_0(4 -x) \frac{dx}{x + s}.$$
Adding and subtracting the first $N+1$ terms in the power series in (\ref{N0near4}), we obtain
\begin{align}\label{intN0addsubtract}
V(t_\perp) = \sum_{j=0}^N a_j \int_0^8  \frac{x^j}{x + s}dx
+ \int_0^8 \bigg(N_0(4-x) -  \sum_{l=0}^N a_l  x^l\bigg)\frac{dx}{x + s}.
\end{align}
For any integer $j \geq 0$, we have
$$\frac{d}{dx} \bigg((-s)^j \ln(x + s) + \sum_{k=0}^{j-1}(-s)^k \frac{x^{j-k}}{j-k}\bigg) = \frac{x^j}{x + s}.$$
Furthermore, 
$$\frac{1}{x+s} = \sum_{j=0}^{N-1} \frac{(-s)^j}{x^{j+1}} + \frac{(-s)^{N}}{x^{N} (s+x)}.$$
Using these identities in (\ref{intN0addsubtract}), we deduce that
\begin{align*}
V(t_\perp)
= &\; \sum_{j=0}^N a_j (-s)^j \ln\Big(\frac{8 + s}{s}\Big) + \sum_{j=0}^N a_j \sum_{k=0}^{j-1}(-s)^k \frac{8^{j-k}}{j-k}
	\\
& + \int_0^8 \bigg(N_0(4-x) -  \sum_{l=0}^N a_l  x^l\bigg)\bigg(\sum_{j=0}^{N-1} \frac{(-s)^j}{x^{j+1}} + \frac{(-s)^{N}}{x^{N} (s+x)}\bigg) dx
\end{align*}
for $s = t_\perp - 4 >0$.
In the first sum, we use the expansion
$$\ln(8+s) = \ln{8} - \sum_{k=1}^N \frac{(-s)^k}{8^k k} + O(s^{N+1}) \qquad \text{as $s \downarrow 0$},$$ 
and in the double sum involving $8^{j-k}$, we change the order of summation. Also, by Lemma \ref{N0near4lemma}, $N_0(4-x) -  \sum_{l=0}^N a_l  x^l = O(x^{N+1})$ as $x \downarrow 0$, so we can split the last integral to find
\begin{align*}
V(t_\perp) = & - \sum_{j=0}^N a_j (-s)^j \ln{s} + \sum_{j=0}^N a_j (-s)^j \bigg(\ln{8} - \sum_{k=1}^N \frac{(-s)^k}{8^k k}\bigg)
+ \sum_{k=0}^{N-1} \sum_{j=k+1}^N a_j (-s)^k \frac{8^{j-k}}{j-k}
	\\
& + \sum_{j=0}^{N-1} \int_0^8 \bigg(N_0(4-x) -  \sum_{l=0}^N a_l  x^l\bigg) \frac{1}{x^{j+1}} dx (-s)^j
+ O(s^N).
\end{align*}
Thus, gathering powers of $s$, we conclude that
\begin{align*}
V(t_\perp) = &\; \sum_{j=0}^{N} (b_j(N) - a_j \ln{s}) (-s)^j + O(s^N)
	\\
 =&\; b_0(N) - a_0 \ln{s} + \sum_{j=1}^{N-1} (b_j(N) - a_j \ln{s}) (-s)^j + O(s^N \ln{s}),
\end{align*}
where the coefficients $a_j$ and $b_j(N)$ are given by (\ref{ajdef}) and (\ref{bjdef}), from which assertion $(\ref{Ucat4item})$ follows. 
\end{proof}

\section{Antiferromagnetic gap near the critical curve: proof of Theorem \ref{criticalth}}\label{criticalsec}
Since $N_0(\epsilon)$ is nonzero only for $|\epsilon| \leq 4$, the zero temperature equation (\ref{DeltaAFat0}) for $\Delta_{\AF}$ implies that
$$\frac{1}{U} = \int_0^\infty  \frac{N_{t_\perp}(\epsilon)}{\sqrt{\Delta_{\AF}(U, t_{\perp}, 0)^2 + \epsilon^2}}  d\epsilon = \frac{1}{2}\int_{t_\perp-4}^{t_\perp+4} \frac{N_0(\epsilon - t_\perp)}{\sqrt{\Delta_{\AF}(U, t_{\perp}, 0)^2 + \epsilon^2}}  d\epsilon$$
for $U > 0$ and $t_\perp \in (4, t_{\perp,c}(U))$.
Changing variables $\epsilon \to \epsilon + t_\perp$ and introducing the notation $f(t_\perp; U) := \Delta_{\AF}(U, t_{\perp}, 0)^2$, we find
\begin{align}\label{N0feq}
\frac{2}{U} = \int_{-4}^{4} \frac{N_0(\epsilon)}{\sqrt{f(t_{\perp}; U) + (\epsilon + t_\perp)^2}}  d\epsilon \qquad \text{for $U > 0$ and $t_\perp \in (4, t_{\perp,c}(U))$.}
\end{align}
Differentiating (\ref{N0feq}) with respect to $t_\perp$, we obtain the equation
$$0 = \int_{-4}^4 N_0(\epsilon) \frac{f'(t_{\perp}; U)+2(\epsilon+t_{\perp})}{2(f(t_{\perp}; U) + (\epsilon + t_\perp)^2)^{3/2}}  d\epsilon,$$
which can be solved for $f'(t_{\perp}; U)$ to yield
$$f'(t_{\perp}; U) = - \frac{\int_{-4}^4 N_0(\epsilon) \frac{\epsilon+t_{\perp}}{(f(t_{\perp}; U) + (\epsilon + t_\perp)^2)^{3/2}}  d\epsilon}{\int_{-4}^4 N_0(\epsilon) \frac{1}{2(f(t_{\perp}; U) + (\epsilon + t_\perp)^2)^{3/2}}  d\epsilon} \qquad \text{for $U > 0$ and $t_\perp \in (4, t_{\perp,c}(U))$},$$
where prime denotes differentiation with respect to $t_\perp$.
Taking the limit $t_{\perp} \uparrow t_{\perp,c}(U)$, and using that $f(t_{\perp,c}(U); U) = 0$, we see that 
$$f'(t_{\perp,c}(U); U) := \lim_{t_{\perp} \uparrow t_{\perp,c}(U)} f'(t_\perp; U)$$
exists and equals
\begin{align}\label{fprimetperpc}
f'(t_{\perp,c}(U); U) =
- \frac{\int_{-4}^4 \frac{ N_0(\epsilon)}{(\epsilon + t_{\perp,c}(U))^2}  d\epsilon}{\int_{-4}^4 \frac{ N_0(\epsilon)}{2(\epsilon + t_{\perp,c}(U))^3}  d\epsilon} < 0.
\end{align}

Similarly, by taking the second derivative of (\ref{N0feq}), solving the resulting equation for $f''(t_{\perp}; U)$, and then letting $t_{\perp} \uparrow t_{\perp,c}(U)$, we see that the limit 
$$f''(t_{\perp,c}(U); U) := \lim_{t_{\perp} \uparrow t_{\perp,c}(U)} f''(t_\perp; U)$$
exists and equals
\begin{align}\label{fprimeprimetperpc}
f''(t_{\perp,c}(U); U) = 2 \frac{\int_{-4}^4 N_0(\epsilon) \big(\frac{3f'(t_{\perp,c}(U); U)^2}{8(\epsilon + t_{\perp,c}(U))^5}  + \frac{3f'(t_{\perp,c}(U); U)}{2(\epsilon + t_{\perp,c}(U))^4} + \frac{1}{(\epsilon + t_{\perp,c}(U))^3} \big) d\epsilon}{\int_{-4}^4  \frac{N_0(\epsilon)}{2(\epsilon + t_{\perp,c}(U))^3}  d\epsilon}.
\end{align}
Note that $f'(t_{\perp,c}(U); U)$ and $f''(t_{\perp,c}(U); U)$ depend smoothly on $U > 0$. 

More generally, suppose $f^{(j)}(t_{\perp,c}(U); U) := \lim_{t_{\perp} \uparrow t_{\perp,c}(U)} f^{(j)}(t_\perp; U)$ exists for each $U > 0$ and depends smoothly on $U > 0$ for each $j = 1, \dots, N-1$. 
The $N$th derivative of (\ref{N0feq}) involves the term $f^{(N)}(t_{\perp}; U) \int_{-4}^4  \frac{N_0(\epsilon)}{2(f(t_{\perp}; U) + (\epsilon + t_\perp)^2)^{3/2}}  d\epsilon$ together with terms involving the lower order derivatives $f^{(j)}(t_{\perp}; U)$ for $j = 0, 1, \dots, N-1$. We can therefore solve for $f^{(N)}(t_{\perp}; U)$ in terms of $\{f^{(j)}(t_{\perp}; U)\}_{j=0}^{N-1}$ and then take the limit $t_{\perp} \uparrow t_{\perp,c}(U)$ to conclude that $f^{(N)}(t_{\perp,c}(U); U) := \lim_{t_{\perp} \uparrow t_{\perp,c}(U)} f^{(N)}(t_{\perp}; U)$ exists for each $U > 0$ and depends smoothly on $U > 0$. 
By induction, $f(t_\perp;U)$ has a smooth extension to $t_\perp \in (4, t_{\perp,c}(U)]$ for each $U > 0$ such that $U \mapsto f^{(N)}(t_{\perp,c}(U); U)$ is a smooth function of $U > 0$ for each integer $N \geq 1$.
Setting 
$$d_j(U) := (-1)^j \frac{f^{(j)}(t_{\perp,c}(U); U)}{j!}, \qquad j = 1, 2, \dots,$$
we obtain (\ref{Deltanearcritical}) from the Taylor expansion of $t_\perp \mapsto f(t_\perp; U)$ at $t_{\perp,c}(U)$.

The explicit expressions for the coefficients $d_1(U)$ and $d_2(U)$ in (\ref{djexplicit}) follow immediately from (\ref{fprimetperpc}) and (\ref{fprimeprimetperpc}), and the expressions in (\ref{djexplicit}) for $d_j(U)$ for $j = 3,4,5$ follow by similar computations. This completes the proof of Theorem  \ref{criticalth}.

\section{N\'eel temperature: proof of Theorem \ref{3DNeelth}}\label{Neelsec}
In this section, we study the asymptotic behavior of the N\'eel temperature $T_N(U, t_{\perp})$ as $U \downarrow 0$ for $t_{\perp} \in (0,4)$. In particular, we prove Theorem \ref{3DNeelth}.

Let $K$ be a compact subset of $(0,4)$ and suppose that $t_{\perp} \in K$.
Adding and subtracting terms, we can write equation (\ref{TNeelequation}) for $T_N = T_N(U,t_{\perp})$ as
\begin{align}\label{Tneeleq2}
\frac{2}{U} - 2 b_0(t_{\perp}) = &\; J_1(t_{\perp}, T_N) + N_0(t_{\perp})J_2(t_{\perp}, T_N) + J_3(t_{\perp}, T_N),
\end{align}
where $b_0(t_{\perp})$ is the function in (\ref{b0tzdef}), and $J_1$, $J_2$, and $J_3$ are defined by
\begin{align*}
& J_1(t_{\perp}, T) = \int_0^4 N_0(\epsilon)  \frac{\tanh(\frac{|\epsilon + t_{\perp}|}{2T}) - 1}{|\epsilon + t_{\perp}|}  d\epsilon,
	\\
& J_2(t_{\perp}, T) = \int_0^4  \frac{\tanh(\frac{|\epsilon - t_{\perp}|}{2T})}{|\epsilon - t_{\perp}|} d\epsilon,
	\\
& J_3(t_{\perp}, T) = \int_0^4 \big(N_0(\epsilon) - N_0(t_{\perp})\big)  \frac{\tanh(\frac{|\epsilon - t_{\perp}|}{2T}) - 1}{|\epsilon - t_{\perp}|} d\epsilon.
\end{align*}

The following three lemmas describe the behavior of $J_1(t_{\perp}, T)$, $J_2(t_{\perp}, T)$, and $J_3(t_{\perp},T)$ for small $T$.

\begin{lemma}\label{J1lemma}
There exist constants $C, c > 0$ such that
$$|J_1(t_{\perp}, T)| \leq C e^{-\frac{c}{T}}  \qquad \text{for all $t_{\perp} \in K$ and $T \in (0,1)$.}$$
\end{lemma}
\begin{proof}
Since $|\tanh(a) - 1 | \leq 2 e^{-2a}$ for $a \geq 0$, we obtain, for $t_{\perp} \in K$ and $T \in (0,1)$,
\begin{align*}
|J_1(t_{\perp}, T)| 
& 
\leq \int_{0}^4 N_0(\epsilon) \frac{2e^{-\frac{\epsilon + t_{\perp}}{T}}}{\epsilon + t_{\perp}} d\epsilon
\leq \frac{2e^{-\frac{t_{\perp}}{T}}}{t_{\perp}}  \int_{0}^4 N_0(\epsilon) d\epsilon
= \frac{e^{-\frac{t_{\perp}}{T}}}{t_{\perp}} 
\leq C e^{-\frac{c}{T}},
\end{align*}
which is the desired conclusion.
\end{proof}

\begin{lemma}\label{J2lemma}
There exist constants $C > 0$ and $\delta \in (0,1)$ such that
$$\bigg|J_2(t_{\perp},T) - \ln\Big(\frac{t_{\perp}}{2T}\Big) - \ln\Big(\frac{4-t_{\perp}}{2T}\Big)  + 2 \ln \Big(\frac{\pi }{4}\Big)- 2\gamma\bigg|
 \leq  C \ln\Big(\frac{1}{T}\Big) e^{-\frac{c}{T}}$$
 for all $t_{\perp} \in K$ and $T \in (0,\delta)$.
\end{lemma}
\begin{proof}
The change of variables $x = \frac{\epsilon - t_{\perp}}{2T}$ yields
$$J_2(t_{\perp},T) = \int_{\frac{- t_{\perp}}{2T}}^{\frac{4 - t_{\perp}}{2T}}  \frac{\tanh{x}}{x} dx
= \varphi\Big(\frac{4 - t_{\perp}}{2T}\Big) + \varphi\Big(\frac{t_{\perp}}{2T}\Big) ,$$
where
$$\varphi(b) = \int_0^b  \frac{\tanh{x}}{x} dx.$$
An integration by parts gives
$$\varphi(b) =  \ln(x) \tanh{x} \Big|_0^b
- \int_0^b \frac{\ln{x}}{\cosh^2{x}}dx
=  \ln(b) \tanh(b)
- \int_{0}^\infty \frac{\ln{x}}{\cosh^2{x}} dx
+ \int_b^\infty  \frac{\ln{x}}{\cosh^2{x}}dx$$
for $t_{\perp} \in K$ and $T \in (0,1)$.
Since $\int_{0}^\infty \frac{\ln{x}}{\cosh^2{x}} dx = \ln \left(\frac{\pi }{4}\right)-\gamma$, it follows that
\begin{align*}
& J_2(t_{\perp},T) - \ln\Big(\frac{t_{\perp}}{2T}\Big) - \ln\Big(\frac{4-t_{\perp}}{2T}\Big)  + 2 \ln \Big(\frac{\pi }{4}\Big)- 2\gamma
= R_1\Big(\frac{t_{\perp}}{2T}\Big) + R_1\Big(\frac{4-t_{\perp}}{2T}\Big),
\end{align*}
where $R_1$ is defined by
\begin{align}\label{R1def}
R_1(a) :=  \ln(a) (\tanh(a) -1)
+ \int_{a}^\infty  \frac{\ln{x}}{\cosh^2{x}}dx.
\end{align}
Since, for $a \geq 2$,
\begin{align}\nonumber
|R_1(a)| 
& \leq \ln(a) |\tanh(a) -1|
+ \int_{a}^\infty  \frac{\ln{x}}{\cosh^2{x}}dx
\leq \ln(a) 2 e^{-2a}
+ C \int_{a}^\infty e^{-2x} \ln(x) dx
	\\ \label{R1logtanh}
& \leq \ln(a) 2 e^{-2a}
+ C e^{-2a} \ln(a) + C\int_{a}^\infty \frac{e^{-2x}}{x}dx
\leq C \ln(a) e^{-2a},
\end{align}
the desired conclusion holds if we choose $\delta \in (0,1)$ such that $\min(\frac{t_{\perp}}{2T}, \frac{4-t_{\perp}}{2T}) \geq 2$ for $t_\perp \in K$ and $T \in (0,\delta)$.
\end{proof}

\begin{lemma}\label{J3lemma}
There exists a constant $C > 0$ such that
$$|J_3(t_{\perp}, T)| \leq  C T  \qquad \text{for all $t_{\perp} \in K$ and $T \in (0,1)$.}$$
\end{lemma}
\begin{proof}
We have
\begin{align*}
N_0(\epsilon) - N_0(t_{\perp})
& = O(\epsilon - t_{\perp}) \qquad \text{as $\epsilon \to 0$}
\end{align*}
uniformly for $t_{\perp} \in K$. Thus we can estimate
\begin{align*}
|J_3(t_{\perp}, T)| \leq C \int_0^4 |\epsilon -t_{\perp}| \frac{|\tanh(\frac{|\epsilon - t_{\perp}|}{2T}) - 1|}{|\epsilon - t_{\perp}|} d\epsilon
\end{align*}
for $t_{\perp} \in K$ and $T \in (0, 1)$.
Using that $|\tanh(a) - 1 | \leq 2 e^{-2a}$ for $a \geq 0$ and then changing variables to $x = \frac{2(\epsilon - t_{\perp})}{T}$, we find, again for $t_{\perp} \in K$ and $T \in (0,1)$,
\begin{align*}
|J_3(t_{\perp}, T)| 
& \leq C \int_{0}^4 e^{-\frac{|\epsilon -t_{\perp}|}{T}} d\epsilon
\leq C T \int_{\R} |x| e^{-\frac{|x|}{2}} dx \leq C T,
\end{align*}
which is the desired conclusion.
\end{proof}

Let $\delta \in (0,1)$ be as in Lemma \ref{J2lemma}.
By Lemma \ref{TNlemma}, $T_N \downarrow 0$ as  $U \downarrow 0$ uniformly for $t_{\perp} \in [0,4]$, so for all sufficiently small $U$ and all $t_\perp \in K$, we have $T_N \in (0,\delta)$. 
We rewrite the equation (\ref{Tneeleq2}) for $T_N$ as follows:
\begin{align}\nonumber
& \frac{2}{U} - 2b_0(t_{\perp})
+ N_0(t_{\perp})\bigg(- \ln\Big(\frac{t_{\perp}}{2T}\Big) - \ln\Big(\frac{4-t_{\perp}}{2T}\Big)  + 2 \ln \Big(\frac{\pi }{4}\Big)- 2\gamma\bigg)
= J_1(t_{\perp}, T_N) 
\\ \nonumber
& \hspace{1cm}+  N_0(t_{\perp})\bigg(J_2(t_{\perp},T) - \ln\Big(\frac{t_{\perp}}{2T}\Big) - \ln\Big(\frac{4-t_{\perp}}{2T}\Big)  + 2 \ln \Big(\frac{\pi }{4}\Big)- 2\gamma\bigg) +  J_3(t_{\perp}, T_N).
\end{align}
Applying Lemmas \ref{J1lemma}--\ref{J3lemma}, we see that the absolute value of the right-hand side is bounded above by
\begin{align*}
&  C \bigg( T_N + \ln\Big(\frac{1}{T_N}\Big) e^{-\frac{c}{T_N}} \bigg)
\leq C T_N
\end{align*}
for $t_{\perp} \in K$ and all sufficiently small $U$. It follows that
\begin{align}\label{logTNR4}
& \frac{2}{U} - 2b_0(t_{\perp})
+ N_0(t_{\perp})
\bigg(- \ln\Big(\frac{t_{\perp}}{2T_N}\Big) - \ln\Big(\frac{4-t_{\perp}}{2T_N}\Big)  + 2 \ln \Big(\frac{\pi }{4}\Big)- 2\gamma\bigg) = R_4(t_{\perp}, T_N),
 \end{align}
 where the function $R_4$ satisfies
\begin{align}\label{R4bound}
|R_4(t_{\perp}, T)| \leq C T \qquad \text{for all $t_{\perp} \in K$ and $T \in (0,\delta)$.}
\end{align}
In particular, $R_4(t_{\perp}, T_N(U,t_{\perp})) = O(1)$ as $U \downarrow 0$ uniformly for $t_{\perp} \in K$.
Solving (\ref{logTNR4}) for $\ln T_N$, we obtain
\begin{align}\label{logTNtzsolvedfor}
\ln T_N(U,t_{\perp})
= \gamma + \ln\Big(\frac{\sqrt{4t_{\perp}(4-t_{\perp})}}{\pi}\Big) - \frac{1}{N_0(t_{\perp})}\bigg(\frac{1}{U} - b_0(t_{\perp})\bigg) + \frac{R_4(t_{\perp}, T_N(U,t_{\perp}))}{2N_0(t_{\perp})},
\end{align}
which implies that $T_N(U,t_{\perp}) = O(e^{-\frac{1}{N_0(t_{\perp}) U}})$ as $U \downarrow 0$ uniformly for $t_{\perp} \in K$. Substituting this into (\ref{R4bound}), we infer that $R_4(t_{\perp}, T_N(U,t_{\perp})) = O(e^{-\frac{1}{N_0(t_{\perp}) U}})$ as $U \downarrow 0$ uniformly for $t_{\perp} \in K$. Inserting this in (\ref{logTNtzsolvedfor}), we arrive at (\ref{TNexpansionbilayer}). This completes the proof of Theorem \ref{3DNeelth}.

\section{Antiferromagnetic gap ratio: proof of Theorem \ref{3Dmeanfieldth}}\label{hatmsec}
In this section, we study the asymptotic behavior of the mean-field ratio $\hat{m}$ defined in (\ref{hatmdef}) as $U \downarrow 0$ for $t_\perp \in (0,4)$. Taking the difference of (\ref{DeltaAFeq}) divided by $\Delta_{\AF}$ and (\ref{TNeelequation}), we obtain
$$0 = \int_0^\infty N_{t_\perp}(\epsilon) \bigg(\frac{\tanh(\frac{\sqrt{\Delta_{\AF}^2 + \epsilon^2}}{2T})}{\sqrt{\Delta_{\AF}^2 + \epsilon^2}}  
- \frac{\tanh(\frac{\epsilon}{2T_N})}{\epsilon}  \bigg)
d\epsilon,
$$
where $\Delta_{\AF} = \Delta_{\AF}(U, t_{\perp}, T)$ and $T_N = T_N(U,t_{\perp})$. Performing the change of variables $\epsilon \to T_N\epsilon$, we get
\begin{align}\label{mAFeq3}
\int_0^\infty N_{t_\perp}(T_N\epsilon)  
\bigg(\frac{\tanh(\frac{\sqrt{\hat{m}^2 + \epsilon^2}}{2T/T_N})}{\sqrt{\hat{m}^2 + \epsilon^2}}  -
\frac{\tanh(\frac{\epsilon}{2})}{\epsilon} \bigg) d\epsilon 
= 0,
\end{align}
where $\hat{m} = \hat{m}(U, t_{\perp}, T)$ is given by (\ref{hatmdef}). We will prove Theorem \ref{3Dmeanfieldth} by approximating $N_{t_\perp}(T_N\epsilon)$ in (\ref{mAFeq3}) by its leading asymptotics for small $T_N\epsilon$.

\subsection{Preliminary lemmas}
We collect four lemmas that are needed for the proof of Theorem \ref{3Dmeanfieldth}.
Recall that, by convention, $\tanh(\cdot/y) = 1$ for $y = 0$.

For any fixed $t_{\perp} \in (0,4)$, the function $\epsilon \mapsto N_{t_\perp}(T_N\epsilon)$ is non-negative for all $\epsilon > 0$ and strictly positive for all sufficiently small $\epsilon > 0$. The next lemma, which establishes an upper bound on $\hat{m}$, therefore follows from (\ref{mAFeq3}) in the same way as \cite[Lemma 6.1]{LL2025}.

\begin{lemma}\label{mhatboundlemma}
For $U> 0$, $t_{\perp} \in (0,4)$, and $T \in [0, T_N(U,t_{\perp})]$, it holds that $\hat{m}(U, t_{\perp}, T) \in [0, 2)$.
\end{lemma}

\begin{lemma}\label{tanhlemma}
There is a constant $C > 0$ such that
\begin{align}\label{tanhdifference}
\bigg|\frac{\tanh(\frac{\sqrt{x^2 + \epsilon^2}}{2y})}{\sqrt{x^2 + \epsilon^2}}  -
\frac{\tanh(\frac{\epsilon}{2})}{\epsilon} \bigg| \leq \frac{C}{\epsilon^3}
\end{align}
for all $x \in [0,2]$, $y \in [0,1]$, and $\epsilon \in [2, +\infty)$. 
\end{lemma}
\begin{proof}
Let $H(x,y,\epsilon)$ denote the left-hand side of (\ref{tanhdifference}). 
The triangle inequality implies that 
\begin{align*}
H(x,y,\epsilon) \leq \bigg|\frac{\tanh(\frac{\sqrt{x^2 + \epsilon^2}}{2y}) - 1}{\sqrt{x^2 + \epsilon^2}}\bigg|  +
\bigg|\frac{\tanh(\frac{\epsilon}{2}) - 1}{\epsilon}\bigg| + 
\frac{1}{\epsilon} \bigg|\frac{1}{\sqrt{(x/\epsilon)^2 + 1}} - 1 \bigg|.
\end{align*}
Since $|\tanh(a) - 1 | \leq 2 e^{-2a}$ for $a \geq 0$ and $|\frac{1}{\sqrt{a + 1}} - 1| \leq \frac{a}{2}$ for $a \geq 0$, we get
\begin{align}\nonumber
H(x,y,\epsilon) & \leq \frac{2 e^{-\frac{\sqrt{x^2 + \epsilon^2}}{y}}}{\sqrt{x^2 + \epsilon^2}} +
\frac{2 e^{-\epsilon}}{\epsilon} + 
 \frac{x^2}{2\epsilon^3} 
 \leq \frac{4 e^{-\epsilon}}{\epsilon} + \frac{x^2}{2\epsilon^3}  \leq \frac{C}{\epsilon^3}
\end{align}
for all $x \in [0,2]$, $y \in [0,1]$, and $\epsilon \in [2, +\infty)$. 
\end{proof}

\begin{lemma}{\upshape \cite[Lemma 6.2]{LL2025}} \label{int02lemma}
$(a)$ For any $r > 0$, there is a constant $C > 0$ such that
\begin{align}
\int_{0}^{2} (1+|\ln{\epsilon}|) \frac{\tanh(\frac{\sqrt{x^2 + \epsilon^2}}{2y})}{\sqrt{x^2 + \epsilon^2}} d\epsilon \leq C
\end{align}
for all $x \in [0,2]$ and $y \in [0,1]$ such that $\sqrt{x^2 + y^2} \geq r$. 

$(b)$ There exist constants $c > 0$ and $r > 0$ such that
\begin{align}
\int_{0}^{2} \frac{\tanh(\frac{\sqrt{x^2 + \epsilon^2}}{2y})}{\sqrt{x^2 + \epsilon^2}} d\epsilon \geq c|\ln(x^2+y^2)|
\end{align}
for all $x \in [0,2]$ and $y \in [0,1]$ such that $0 < \sqrt{x^2 + y^2} \leq r$. 
\end{lemma}

\begin{lemma}{\upshape \cite[Lemma 6.3 with $\delta = 0$]{LL2025}}\label{intxylemma}
$(a)$ There exists a $C > 0$ such that
\begin{align}\label{intepsilonxyz}
\int_{z}^{\infty}  \bigg|\frac{\tanh(\frac{\sqrt{x^2 + \epsilon^2}}{2y})}{\sqrt{x^2 + \epsilon^2}}  -
\frac{\tanh(\frac{\epsilon}{2})}{\epsilon} \bigg| d\epsilon \leq C \bigg(\frac{e^{-z}}{z}  + \frac{1}{z^2}\bigg),
\end{align}
for  $x \in [0,2]$, $y \in [0,1]$, and $z \geq 2$.

$(b)$ There exists a $C > 0$ such that
\begin{align}\label{intepsilon2xyz}
\int_{2}^{z} \epsilon^2 
\bigg|\frac{\tanh(\frac{\sqrt{x^2 + \epsilon^2}}{2y})}{\sqrt{x^2 + \epsilon^2}}  -
\frac{\tanh(\frac{\epsilon}{2})}{\epsilon} \bigg| d\epsilon \leq C +  C|\ln z |
\end{align}
for $x \in [0,2]$, $y \in [0,1]$, and $z \geq 2$. 
\end{lemma}

\subsection{Proof of Theorem \ref{3Dmeanfieldth}}\label{3Dmeanfieldsubsec}
Assume that $t_{\perp} \in K_1$ and $T/T_N(U, t_{\perp}) \in K_2$, where $K_1$ and $K_2$ are given compact subsets of $(0, 4)$ and $[0,1)$, respectively.

Since $N_{t_\perp}(\epsilon) = 0$ for $\epsilon > 4 + t_\perp$, we can write (\ref{mAFeq3}) as
\begin{align}\label{mAFeq43D}
 \int_0^{\frac{4 + t_{\perp}}{T_N}} 
\bigg(\frac{\tanh(\frac{\sqrt{\hat{m}^2 + \epsilon^2}}{2T/T_N})}{\sqrt{\hat{m}^2 + \epsilon^2}}  -
\frac{\tanh(\frac{\epsilon}{2})}{\epsilon} \bigg) d\epsilon 
= E_1(U,t_{\perp},T),
\end{align}
where $\hat{m} = \hat{m}(U, t_{\perp}, T)$, $T_N = T_N(U,t_{\perp})$, and
\begin{align}\label{E1def}
E_1(U,t_{\perp},T)
:= -\int_0^{\frac{4 + t_{\perp}}{T_N}} \frac{N_{t_\perp}(T_N\epsilon)  - N_{t_\perp}(0)}{N_{t_\perp}(0)} 
\bigg(\frac{\tanh(\frac{\sqrt{\hat{m}^2 + \epsilon^2}}{2T/T_N})}{\sqrt{\hat{m}^2 + \epsilon^2}}  -
\frac{\tanh(\frac{\epsilon}{2})}{\epsilon} \bigg) d\epsilon.
\end{align}
Equation (\ref{mAFeq43D}) implies that
\begin{align}\label{mAFeq6}
& J(\hat{m}, T/T_N)
= E_1(U,t_{\perp},T) + E_2(U,t_{\perp},T),
\end{align}
where $J$ is the function defined in (\ref{Jdef}) and
$$E_2(U,t_{\perp},T) := \int_{\frac{4 + t_{\perp}}{T_N}}^\infty \bigg(\frac{\tanh(\frac{\sqrt{\hat{m}^2 + \epsilon^2}}{2T/T_N})}{\sqrt{\hat{m}^2 + \epsilon^2}}  -
\frac{\tanh(\frac{\epsilon}{2})}{\epsilon} \bigg) d\epsilon.$$
Our next goal is to show that $J(\hat{m}, T/T_N)$ is small for small $U$. Since $f_{\BCS}(y)$ is defined as the unique solution of $J(f_{\BCS}(y), y) =0$, this can be used to show that $\hat{m}(U,t_\perp,T)$ tends to $f_{\BCS}(T/T_N)$ as $U \downarrow 0$. To complete these steps, we need a few lemmas. The first lemma estimates the function $E_2$.

\begin{lemma}\label{E2lemma}
There is a constant $C > 0$ such that
$$|E_2(U,t_{\perp},T)| \leq Ce^{- \frac{2}{N_{t_\perp}(0) U}}$$
for all sufficiently small $U> 0$, all $t_{\perp} \in K_1$, and all $T$ such that $T/T_N(U,t_\perp) \in K_2$.
\end{lemma}
\begin{proof}
By Theorem \ref{3DNeelth},
\begin{align}\label{3DTNbound}
T_N(U, t_{\perp}) = O\big(e^{- \frac{1}{N_{t_\perp}(0) U}}\big)
\qquad \text{as $U \downarrow 0$}
\end{align}
uniformly for $t_{\perp} \in K_1$. 
Since $N_{t_\perp}(0) \leq C$ for all $t_{\perp} \in K_1$, we see that $T_N(U, t_{\perp}) \to 0$ as $U \downarrow 0$ uniformly for $t_{\perp} \in K_1$. Furthermore, by Lemma \ref{mhatboundlemma}, $\hat{m} \in [0, 2)$ for $t_{\perp} \in K_1$ and $T/T_N \in K_2$. 
Thus Lemma \ref{intxylemma}$(a)$ yields
\begin{align*}
|E_2(U,t_{\perp},T)| & \leq \int_{\frac{4 + t_{\perp}}{T_N}}^\infty \bigg|\frac{\tanh(\frac{\sqrt{\hat{m}^2 + \epsilon^2}}{2T/T_N})}{\sqrt{\hat{m}^2 + \epsilon^2}}  -
\frac{\tanh(\frac{\epsilon}{2})}{\epsilon} \bigg| d\epsilon 
\leq C\big(T_N e^{-\frac{4}{T_N}}  + T_N^2 \big)
	\\
& \leq C T_N^2  \leq Ce^{- \frac{2}{N_{t_\perp}(0) U}}
\end{align*}
for all sufficiently small $U> 0$, uniformly for $t_{\perp} \in K_1$ and $T/T_N \in K_2$.
\end{proof}

The next lemma provides an upper bound for the function 
\begin{align}\label{Btperpxytaudef}
B(t_{\perp}, x,y,\tau) := \int_2^\infty N_{t_\perp}(\tau \epsilon)  
\bigg|\frac{\tanh(\frac{\sqrt{x^2 + \epsilon^2}}{2y})}{\sqrt{x^2 + \epsilon^2}}  -
\frac{\tanh(\frac{\epsilon}{2})}{\epsilon} \bigg| d\epsilon \geq 0.
\end{align}
We let $\delta > 0$ denote the constant
\begin{align}\label{deltadef}
\delta := \frac{1}{4}\min_{t_{\perp} \in K_1} t_\perp.
\end{align}

\begin{lemma}\label{Blemma}
There is a constant $C > 0$ such that
\begin{align}\label{Bbound}
B(t_{\perp}, x,y,\tau) \leq C
\end{align}
for all $t_\perp \in K_1$, $\tau \in [0,\delta]$, $x \in [0,2]$, and $y \in [0,1]$.
\end{lemma}
\begin{proof}
For $t_\perp \in K_1$ and $\epsilon \geq 0$, we have
$$N_{t_\perp}(\epsilon) \leq C\big( 1 + |\ln|\epsilon -t_\perp| |\big),$$
and hence, using also Lemma \ref{tanhlemma},
\begin{align*}
B(t_{\perp}, x,y,\tau) \leq&\; C \int_2^\infty \big( 1 + |\ln|\tau \epsilon -t_\perp| |\big)  
\bigg|\frac{\tanh(\frac{\sqrt{x^2 + \epsilon^2}}{2y})}{\sqrt{x^2 + \epsilon^2}}  -
\frac{\tanh(\frac{\epsilon}{2})}{\epsilon} \bigg| d\epsilon
	\\
\leq &\; C + C\int_{2}^{\infty}\frac{|\ln|\tau \epsilon -t_\perp| |}{\epsilon^3} d\epsilon
\end{align*}
for $t_\perp \in K_1$, $\tau \in [0,\delta]$, $x \in [0,2]$, and $y \in [0,1]$.
If $\tau = 0$, then clearly $B(t_{\perp}, x,y,\tau) \leq C$. If $\tau > 0$, then, using the change of variables $\epsilon = \alpha/\tau$ and the fact that $t_\perp$ remains a bounded distance away from $2\delta$, we find
\begin{align*}
B(t_{\perp}, x,y,\tau) \leq &\; C + C\tau^2 \int_{2\tau}^{\infty}\frac{|\ln|\alpha - t_\perp| |}{\alpha^3} d\alpha
	\\
= &\;  C + C\tau^2 \int_{2\tau}^{2\delta}\frac{|\ln(t_\perp - \alpha) |}{\alpha^3} d\alpha
+ C\tau^2 \int_{2\delta}^\infty \frac{|\ln|\alpha - t_\perp| |}{\alpha^3} d\alpha
	\\
\leq &\;  C + C\tau^2 \int_{2\tau}^{2\delta}\frac{1}{\alpha^3} d\alpha
+ C \tau^2
\leq  C
\end{align*}
for $t_\perp \in K_1$, $\tau \in [0,\delta]$, $x \in [0,2]$, and $y \in [0,1]$. This proves (\ref{Bbound}).
\end{proof}

\begin{lemma}\label{3DhatmTlemma}
There is an $r > 0$ such that
$$\sqrt{\hat{m}(U, t_{\perp}, T)^2 + \Big(\frac{T}{T_N(U,t_{\perp})}\Big)^2} \geq r$$ 
for all sufficiently small $U > 0$, all $t_{\perp} \in K_1$, and all $T$ such that $T/T_N(U, t_{\perp}) \in K_2$.
\end{lemma}
\begin{proof}
The logarithmic singularities of $N_{t_\perp}(\epsilon)$ lie at $\epsilon = \pm t_\perp$. Therefore, there are constants $c_1, c_2 > 0$ such that 
\begin{align}\label{Ntperpc1c2}
c_1 < N_{t_\perp}(\tau \epsilon) < c_2
\end{align}
for all $\tau \in [0,\delta]$, $\epsilon \in [0,2]$, and $t_{\perp} \in K_1$.

By Theorem \ref{3DNeelth}, $T_N(U,t_{\perp}) \downarrow 0$ as $U \downarrow 0$ uniformly for $t_{\perp} \in K_1$.
Hence, in view of (\ref{mAFeq3}), it is sufficient to prove that there is an $r>0$ such that
\begin{align}\label{3Dintxynonzero}
A(t_{\perp}, x,y,\tau) := \int_0^\infty N_{t_\perp}(\tau \epsilon)  
\bigg(\frac{\tanh(\frac{\sqrt{x^2 + \epsilon^2}}{2y})}{\sqrt{x^2 + \epsilon^2}}  -
\frac{\tanh(\frac{\epsilon}{2})}{\epsilon} \bigg) d\epsilon \neq 0
\end{align}
for all $(t_\perp, \tau, x,y) \in S_r$, where
$$S_r :=\{(t_\perp, \tau, x,y) \, | \, \text{$t_{\perp} \in K_1$, $\tau \in [0,\delta]$, $x \geq 0$, $y \geq 0$, $\sqrt{x^2 + y^2} \leq r$}\}.$$
By (\ref{Ntperpc1c2}),
\begin{align*}
A(t_{\perp}, x,y,\tau) \geq &\; c_1 \int_0^2 \frac{\tanh(\frac{\sqrt{x^2 + \epsilon^2}}{2y})}{\sqrt{x^2 + \epsilon^2}} d\epsilon
- c_2 \int_0^2 \frac{\tanh(\frac{\epsilon}{2})}{\epsilon} d\epsilon - B(t_{\perp}, x,y,\tau)
\end{align*}
for all $(t_\perp, \tau, x,y) \in S_r$, where $B(t_{\perp}, x,y,\tau)$ is the function in (\ref{Btperpxytaudef}). By Lemma \ref{Blemma}, $B(t_{\perp}, x,y,\tau)$ remains bounded for $(t_\perp, \tau, x,y) \in S_1$.
Employing also Lemma \ref{int02lemma}$(b)$, we conclude that there are constants $r, c, C_1 > 0$ such that
\begin{align}
A(t_{\perp}, x,y,\tau) \geq c |\ln(x^2+y^2)| - C_1
\end{align}
for all $(t_\perp, \tau, x,y) \in S_{r}$. Since $|\ln(x^2+y^2)|$ can be made arbitrarily large positive by choosing $r > 0$ small, we find that (\ref{3Dintxynonzero}) holds for all $(t_\perp, \tau, x,y) \in S_{r}$ if $r > 0$ is sufficiently small. 
\end{proof}

Using the above lemmas, we can estimate the function $E_1$. 

\begin{lemma}\label{E1lemma}
There is a constant $C > 0$ such that
\begin{align}\label{Eestimate}
|E_1(U,t_{\perp},T)| \leq C U^{-1}e^{- \frac{2}{N_{t_\perp}(0) U}}
\end{align}
for all sufficiently small $U> 0$, all $t_{\perp} \in K_1$, and all $T$ such that $T/T_N \in K_2$.
\end{lemma}
\begin{proof}
The function $N_{t_\perp}(\epsilon)$ has a logarithmic singularity at $\epsilon = t_\perp$ and we have
$$\bigg|\frac{N_{t_\perp}(\epsilon)  - N_{t_\perp}(0)}{N_{t_\perp}(0)} \bigg| \leq  
\begin{cases}
 C \epsilon^2   & \text{for $t_{\perp} \in K_1$ and $\epsilon \in [0, +\infty)$ with $|\epsilon - t_\perp| \geq \delta$},
	\\
C |\ln|\epsilon - t_\perp||  & \text{for $t_{\perp} \in K_1$ and $\epsilon \in [0, +\infty)$ with $|\epsilon - t_\perp| \leq \delta$},
\end{cases}
$$
where we recall that $\delta = \min_{t_{\perp} \in K_1} t_\perp/4 > 0$ was defined in (\ref{deltadef}).
Using these estimates in the definition (\ref{E1def}) of $E_1$, we obtain
\begin{align}\nonumber
|E_1(U,t_{\perp},T)| \leq &\; \int_0^{\frac{4 + t_{\perp}}{T_N}} \bigg| \frac{N_{t_\perp}(T_N\epsilon)  - N_{t_\perp}(0)}{N_{t_\perp}(0)} \bigg|
\bigg|\frac{\tanh(\frac{\sqrt{\hat{m}^2 + \epsilon^2}}{2T/T_N})}{\sqrt{\hat{m}^2 + \epsilon^2}}  -
\frac{\tanh(\frac{\epsilon}{2})}{\epsilon} \bigg| d\epsilon
	\\ \label{E1estimate}
\leq &\; C E_{1,1}(U,t_{\perp},T) + C E_{1,2}(U,t_{\perp},T)
\end{align}
for all sufficiently small $U> 0$, $t_{\perp} \in K_1$, and $T/T_N \in K_2$, where
\begin{align*}
& E_{1,1}(U,t_{\perp},T) := T_N^2 \int_{0}^{\frac{4 + t_{\perp}}{T_N}} \epsilon^2
\bigg|\frac{\tanh(\frac{\sqrt{\hat{m}^2 + \epsilon^2}}{2T/T_N})}{\sqrt{\hat{m}^2 + \epsilon^2}}  -
\frac{\tanh(\frac{\epsilon}{2})}{\epsilon} \bigg| d\epsilon,
		\\
& E_{1,2}(U,t_{\perp},T) := \int_{\frac{t_{\perp} - \delta}{T_N}}^{\frac{t_{\perp} + \delta}{T_N}} |\ln|T_N \epsilon - t_\perp||
\bigg|\frac{\tanh(\frac{\sqrt{\hat{m}^2 + \epsilon^2}}{2T/T_N})}{\sqrt{\hat{m}^2 + \epsilon^2}}  -
\frac{\tanh(\frac{\epsilon}{2})}{\epsilon} \bigg| d\epsilon.
\end{align*}
Since $\hat{m} \in [0,2)$ and $\sqrt{\hat{m}^2 + (T/T_N)^2} \geq r > 0$ by Lemma \ref{mhatboundlemma} and Lemma \ref{3DhatmTlemma}, we may employ Lemma \ref{int02lemma}$(a)$ and Lemma \ref{intxylemma}$(b)$ to get
\begin{align}\label{E11estimate}
E_{1,1}(U,t_{\perp},T) \leq C T_N^2 \big(1 + |\ln T_N | \big)
\leq C T_N^2 |\ln T_N |
\end{align}
for all sufficiently small $U> 0$ and for $t_{\perp} \in K_1$ and $T/T_N \in K_2$.
On the other hand, since $\frac{t_{\perp} - \delta}{T_N} > 2$ for all small enough $U>0$, Lemma \ref{mhatboundlemma} and Lemma \ref{tanhlemma} imply that
\begin{align*}
& E_{1,2}(U,t_{\perp},T) \leq C \int_{\frac{t_{\perp} - \delta}{T_N}}^{\frac{t_{\perp} + \delta}{T_N}} \frac{|\ln|T_N \epsilon - t_\perp||}{\epsilon^3} d\epsilon.
\end{align*}
The change of variables $\epsilon \to \epsilon/T_N $ then yields
\begin{align}\label{E12estimate}
& E_{1,2}(U,t_{\perp},T) \leq C T_N^2 \int_{t_{\perp} - \delta}^{t_{\perp} + \delta} \frac{|\ln|\epsilon - t_\perp||}{\epsilon^3}d\epsilon \leq C T_N^2 
\end{align}
for all sufficiently small $U> 0$, $t_{\perp} \in K_1$, and $T/T_N \in K_2$. The desired conclusion follows from (\ref{E1estimate}), (\ref{E11estimate}), (\ref{E12estimate}), and (\ref{3DTNbound}).
\end{proof}

Inserting the estimates of Lemma \ref{E2lemma} and Lemma \ref{E1lemma} in (\ref{mAFeq6}), we get
\begin{align}\label{Jestimate}
& J(\hat{m}, T/T_N) = O\Big(U^{-1} e^{- \frac{2}{N_{t_\perp}(0) U}} \Big)
\end{align}
as $U \downarrow 0$ uniformly for $t_{\perp} \in K_1$ and $T/T_N \in K_2$.
Furthermore, for all $x \geq 0$ and all $y \in [0,1]$ such that $e^{J(x, y)} y \leq 1$, it holds that (see \cite[Lemma 6.5]{LL2025})
\begin{align}\label{fBCSJrelation}
x = f_{\BCS}(e^{J(x, y)} y) e^{-J(x, y)}.
\end{align}
Therefore,
\begin{align}\label{mhatfBCS}
\hat{m} = f_{\BCS}\Big(e^{J(\hat{m}, T/T_N)}\frac{T}{T_N}\Big) e^{-J(\hat{m}, T/T_N)}.
\end{align}
Since $f_{\BCS}(y)$ and $f_{\BCS}'(y)$ are bounded on any compact subset of $[0, 1)$, Taylor's theorem applied to $z \mapsto f_{\BCS}(e^{z}y)e^{-z}$ shows that
\begin{align}\label{fBCStaylor}
f_{\BCS}(e^{z}y)e^{-z} = &\; f_{\BCS}(y) + O(z) \qquad \text{as $z \to 0$}
\end{align}
uniformly for $y \in K_2$. Thus (\ref{Jestimate}) and (\ref{mhatfBCS}) imply that
\begin{align}\label{mhattaylor}
\hat{m} = &\; f_{\BCS}\Big(\frac{T}{T_N}\Big) + O\Big(U^{-1} e^{- \frac{2}{N_{t_\perp}(0) U}} \Big)
\end{align}
as $U \downarrow 0$ uniformly for $t_{\perp} \in K_1$ and $T$ such that $T/T_N \in K_2$.
This completes the proof of Theorem \ref{3Dmeanfieldth}.

\section{Outlook: Numerical phase diagrams away from half-filling}\label{numericalsec}
In this section, we present results of numerical experiments for the bilayer model away from half-filling, and we discuss future research directions suggested by these experiments.

\begin{figure}
\bigskip
\begin{center}
\hspace{-.5cm}
\begin{overpic}[width=.47\textwidth]{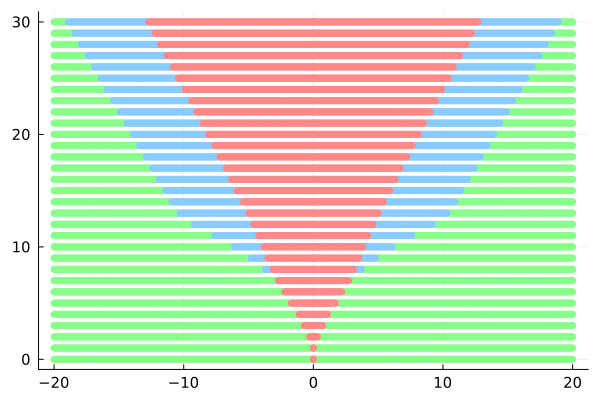}
      \put(40,68){\footnotesize $(a)$ $t_\perp = 0$}
      \put(5,67){\footnotesize $U$}
      \put(100,4.2){\footnotesize $\mu$}
   \end{overpic}
   \hspace{0.3cm}
\begin{overpic}[width=.47\textwidth]{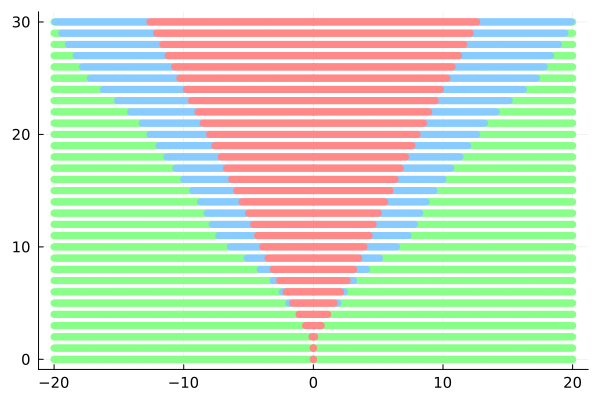}
      \put(40,68){\footnotesize $(b)$ $t_\perp = 1$}
      \put(5,67){\footnotesize $U$}
      \put(100,4.2){\footnotesize $\mu$}
    \end{overpic}
	\vspace{0.8cm}
	\\
\hspace{-.5cm}
\begin{overpic}[width=.47\textwidth]{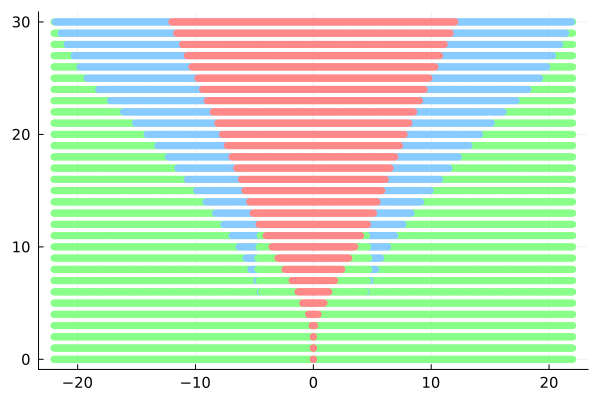}
      \put(40,68){\footnotesize $(c)$ $t_\perp = 3$}
      \put(5,67){\footnotesize $U$}
      \put(100,4.2){\footnotesize $\mu$}
   \end{overpic}
   \hspace{0.3cm}
\begin{overpic}[width=.47\textwidth]{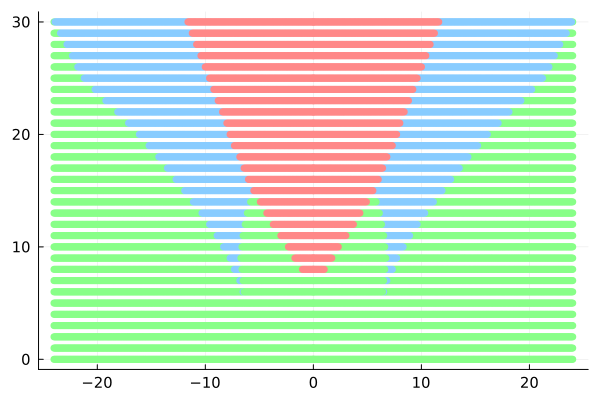}
      \put(40,68){\footnotesize $(d)$ $t_\perp = 5$}
      \put(5,67){\footnotesize $U$}
      \put(100,4.2){\footnotesize $\mu$}
    \end{overpic}
    \begin{figuretext}\label{roughUvsmufig}
       The phase diagram for the bilayer Hubbard model at zero temperature obtained by Hartree--Fock theory restricted to the states in (\ref{dvecmdef}) for (a) $t_\perp = 0$, (b) $t_\perp = 1$, (c) $t_\perp = 3$, and (d) $t_\perp = 5$. 
The phases are shown as a function of the chemical potential $\mu$ and the coupling $U$. 
			AF, F, and P regions are red, blue, and green, respectively. 
      \end{figuretext}
     \end{center}
\end{figure}

\begin{figure}
\bigskip
\begin{center}
\hspace{-.5cm}
\begin{overpic}[width=.47\textwidth]{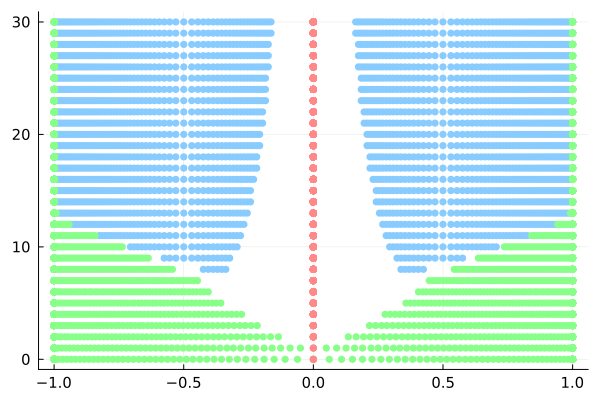}
      \put(40,68){\footnotesize $(a)$ $t_\perp = 0$}
      \put(5,67){\footnotesize $U$}
      \put(100,4.2){\footnotesize $\nu$}
   \end{overpic}
   \hspace{0.3cm}
\begin{overpic}[width=.47\textwidth]{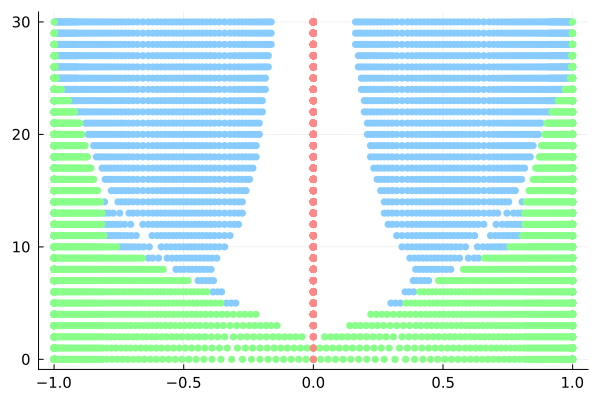}
      \put(40,68){\footnotesize $(b)$ $t_\perp = 1$}
      \put(5,67){\footnotesize $U$}
      \put(100,4.2){\footnotesize $\nu$}
    \end{overpic}
	\vspace{0.8cm}
	\\
\hspace{-.5cm}
\begin{overpic}[width=.47\textwidth]{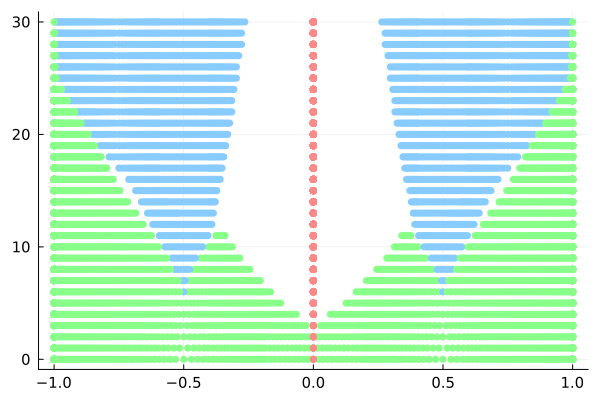}
      \put(40,68){\footnotesize $(c)$ $t_\perp = 3$}
      \put(5,67){\footnotesize $U$}
      \put(100,4.2){\footnotesize $\nu$}
   \end{overpic}
   \hspace{0.3cm}
\begin{overpic}[width=.47\textwidth]{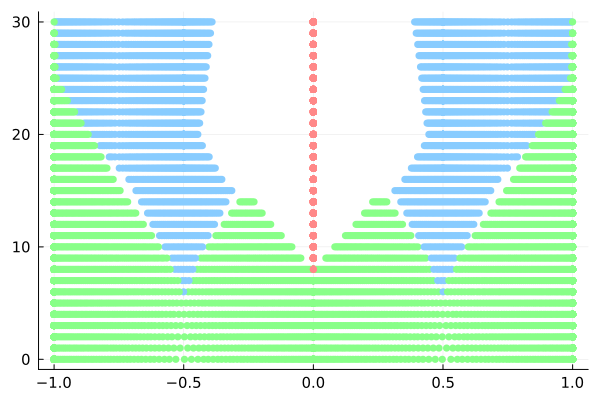}
      \put(40,68){\footnotesize $(d)$ $t_\perp = 5$}
      \put(5,67){\footnotesize $U$}
      \put(100,4.2){\footnotesize $\nu$}
    \end{overpic}
   \begin{figuretext}\label{roughUvsnufig}
       The same phase diagrams as in Figure \ref{roughUvsmufig} but here displayed as a function of doping $\nu$ and coupling $U$. AF, F, and P regions are red, blue, and green, respectively. The white regions indicate mixed phases where new physics can be expected.
      \end{figuretext}
     \end{center}
\end{figure}

Figure \ref{roughUvsmufig} shows numerically computed $U$ vs. $\mu$ phase diagrams for four different values of $t_\perp$. The phase diagrams are obtained using restricted Hartree--Fock theory applied to the family of staggered states in (\ref{dvecmdef}) parametrized by the four parameters $d_0$, $d_1$, $m_0$, and $m_1$. 
This family includes three phases which are known to be relevant for Hubbard-like models \cite{G1963,H1963,K1963,P1966}: the antiferromagnetic (AF) phase, the ferromagnetic (F) phase, and the paramagnetic (P) phase; these phases correspond to the following solutions $(d^*_0,d^*_1,m^*_0,m^*_1)$ of restricted Hartree--Fock theory:
\begin{equation}\label{MF}  
	\begin{split} 
\text{AF}:&\quad (d^*_0,d^*_1,m^*_0,m^*_1) = (d_{\mathrm{AF}},0,0,m_{\mathrm{AF}}), \\	
\text{F}:&\quad (d^*_0,d^*_1,m^*_0,m^*_1)= (d_{\mathrm{F}},0,m_{\mathrm{F}},0), \\
\text{P}:&\quad (d^*_0,d^*_1,m^*_0,m^*_1)= (d_{\mathrm{P}},0,0,0),
	\end{split} 
\end{equation}
with $m_{\mathrm{F}}> 0$ and $m_{\mathrm{AF}}> 0$. The procedure we used to obtain these solutions is summarized in Appendix~\ref{awayfromhalffillingapp}. 
In Figure \ref{roughUvsmufig}, AF, F, and P regions are red, blue, and green, respectively. Figure \ref{roughUvsmufig} suggests that the system displays either AF, F, or P order for any choice of $\mu$ and $U$; more complicated states (such as charge density wave states with $d_1 \neq 0$) appear to be absent.  

Figure \ref{roughUvsnufig} shows the same phase diagrams as Figure \ref{roughUvsmufig} except that the phase is now shown as a function of $U$ vs. $\nu$, where $\nu \in [-1,1]$ is the doping. More precisely, for each dot in the phase diagrams of Figure \ref{roughUvsmufig}, we have computed the corresponding doping $\nu$ and inserted it in the $(\nu, U)$ plane. 

The following two research questions would be interesting to pursue in light of the numerical experiments of Figures \ref{roughUvsmufig} and \ref{roughUvsnufig}.

\begin{enumerate}[$1.$]
\item In \cite{CLL2025}, we used analytic tools to construct the Hartree--Fock phase diagram of the 2D Hubbard model. It would be interesting to implement the method of \cite{CLL2025} also to the bilayer Hubbard model. This would give analytic asymptotic expressions for the phase-separating curves in Figures \ref{roughUvsmufig} and \ref{roughUvsnufig}. Since the density of states $N_{t_\perp}(\epsilon)$ jumps from $1/(8\pi)$ to $0$ at $\epsilon = \pm (4+t_{\perp})$ if $t_{\perp} > 0$ (see Figure \ref{N0Ntperpschematicfig}), we conjecture that the phase boundary between the F and P phases approaches $U = 8\pi \approx 25.13$ as the doping approaches $\pm 1$ for the bilayer model. (If $t_{\perp} = 0$, it is known from \cite{CLL2025} that the phase boundary between the F and P phases approaches $U = 4\pi \approx 12.57$, corresponding to the fact that $N_{0}(\epsilon)$ jumps from $1/(4\pi)$ to $0$ at $\epsilon = \pm 4$.)
This conjecture is consistent with Figure \ref{roughUvsnufig} and is an example of a question that can be settled by implementing the method of \cite{CLL2025}.

\item The phase diagrams of Figure \ref{roughUvsnufig} exhibit empty areas free of dots: most noticeable are the empty areas on either side of the AF phase, but there are also signs of narrow empty areas at or near the interface between the F and P regions.
 If $(\nu, U)$ belongs to one of these areas, no AF, F, or P state exists with the given value of $(\nu, U)$. These empty areas are therefore candidates for so-called mixed phases \cite{LL2026}. If $(\nu, U)$ belongs to an empty area, the only way for our restricted Hartree--Fock approximation to give rise to a state with doping $\nu$ is for the state to be a mixture of two different pure phases. More likely, the restricted Hartree--Fock approximation breaks down in this region, which means that mixed phases serve as indicators for new physics. It is an interesting analytical project to prove the existence of mixed phases for the bilayer model. It is an interesting experimental project to tune a cold atom system simulating the bilayer model to a mixed region and explore what kinds of states will be discovered there---one should find states of a more exotic character than AF, F, or P states.

\end{enumerate}

\appendix

\section{Hartree--Fock theory for the bilayer Hubbard model}\label{derivationapp}
Consider the bilayer Hubbard model defined by the Hamiltonian operator $H =  H_0 + H_1$ in (\ref{H0H1def}) for $t=1$, acting on the fermion Fock space $\mathcal{F}$.
A state $\rho$ is a self-adjoint map $\mathcal{F} \to \mathcal{F}$ with nonnegative eigenvalues and unit trace. 
The most probable state (the equilibrium state) of the system for given values of $U$, $t_\perp$, and $\mu$, is the state that minimizes the grand canonical potential $\Omega(\rho)$ defined by
$$\Omega(\rho) = \Tr_{\mathcal{F}}[\rho H] + T \Tr_{\mathcal{F}}[\rho \ln \rho],$$
where $T\geq 0$ is the temperature and $\Tr_{\mathcal{F}}$ is the trace on $\mathcal{F}$.

\subsection{Unrestricted Hartree--Fock theory}
Let us fix a positive temperature $T>0$. Then the unique absolute minimum of $\Omega$ over the set of all states $\rho$ is the Gibbs state $\rho_{\mathrm{G}}$ given by
\begin{equation}\label{rhoG} 
	\rho_{\mathrm{G}} = \frac{e^{-H/T}}{\Tr_{\mathcal{F}}(e^{-H/T})}.
\end{equation} 
In unrestricted HF theory one aims to find the minimal value $\Omega_{HF}$ of $\Omega$ over all so-called quasi-free states, instead of over all states; this amounts to solving the minimization problem
\begin{align}\label{OmegaHF}
\Omega_{HF} = \min_{(d,\vec{m}) \in D \times M} \Omega(\rho_{d,\vec{m}}),
\end{align}
where the compact sets $D$ and $M$ are defined by
\begin{align*}
& D = \{d:\Lambda_L \times \{1,2\} \to \R \,|\; \text{$|d(\mathbf{x},j)| \leq 1$ for all $(\mathbf{x}, j) \in \Lambda_L \times \{1,2\}$}\},
	\\
& M = \{\vec{m}:\Lambda_L \times \{1,2\} \to \R^3 \,|\; \text{$|\vec{m}(\mathbf{x}, j)| \leq 1$ for all $(\mathbf{x}, j) \in \Lambda_L \times \{1,2\}$}\},
\end{align*}
and the quasi-free state $\rho_{d,\vec{m}}$ associated to $(d,\vec{m}) \in D \times M$ is defined as follows:
\begin{align}
 \rho_{d,\vec{m}} := \frac{e^{-H_{d,\vec{m}}/T}}{\Tr(e^{-H_{d,\vec{m}}/T})}, 
\end{align} 
where
\begin{align}
 H_{d,\vec{m}} := H_0 + \frac{U}{2}\sum_{(\mathbf{x}, j) \in \Lambda_L \times \{1,2\}} \big( d(\mathbf{x},j)(n_{\mathbf{x} j}-1)-\vec{m}(\mathbf{x},j)\cdot \vec{s}_{\mathbf{x} j} 
\big).
\end{align} 
Since the spin operators can be written as $\vec{s}_{\mathbf{x} j} := \sum_{\tau,\tau'} c_{\mathbf{x} j \tau}^\dagger \vec{\sigma}_{\tau,\tau'} c_{\mathbf{x} j \tau'}$ with $\vec{\sigma} = (\sigma_1, \sigma_2, \sigma_3)$ the vector of Pauli matrices, we have  
$$
H_{d,\vec{m}} = \sum_{\mathbf{x},\mathbf{x}'\in\Lambda_L}\sum_{j,j'=1,2}\sum_{\tau,\tau'=\uparrow,\downarrow} h_{\mathbf{x} j \tau, \mathbf{x}' j' \tau'}\big( c_{\mathbf{x}' j' \tau'} c_{\mathbf{x} j \tau}^\dagger- \tfrac12 \delta_{\mathbf{x}\mathbf{x}'} \delta_{j j'} \delta_{\tau \tau'}\big) 
$$
with 
\begin{align}\label{hdef}
h_{\mathbf{x} j \tau, \mathbf{x}' j' \tau'} = t_{\mathbf{x}j \mathbf{x}' j'}  \delta_{\tau \tau'} - \mu\delta_{\mathbf{x}\mathbf{x}'} \delta_{j j'} \delta_{\tau \tau'} + \frac{U}{2}  \delta_{\mathbf{x}\mathbf{x}'} \delta_{j j'} \big(d(\mathbf{x}, j)\delta_{\tau \tau'} - \vec{m}(\mathbf{x}, j) \cdot \vec{\sigma}_{\tau \tau'}\big)
\end{align} 
the matrix elements of a $4L^2 \times 4L^2$-matrix $h$ called the single-particle Hamiltonian. 

Minimizers of (\ref{OmegaHF}) can be found effectively using the single particle Hamiltonian $h$ as follows. Consider the function $G:K \times B \to \R$ defined by
\begin{align}\label{calGdef}
G(d, \vec{m}) = \frac{U}{4} \sum_{\mathbf{x} \in \Lambda_L} \sum_{j=1,2} \big(|\vec{m}(\mathbf{x},j)|^2 - d(\mathbf{x},j)^2\big)
- \Tr_{\mathcal{H}}\big[\Ln_T(h)\big],
\end{align}
where $\Tr_{\mathcal{H}}$ is the trace on the single-particle Hilbert space $\mathcal{H}$ and
$$\Ln_T(E) := T \ln\bigg( 2 \cosh\bigg(\frac{E}{2 T}\bigg)\bigg).$$
If $(d^*,\vec{m}^*) \in D \times M$ is a stationary point of $G$ such that
$$G(d^*,\vec{m}^*) = \min_{\vec{m} \in M}\max_{d \in D} G(d, m),$$
then the quasi-free state $\rho_{d^*,\vec{m}^*}$ is a minimizer of $\Omega$ among all quasi-free states and, in addition, $\Omega(\rho_{d^*,\vec{m}^*}) = G(d^*,\vec{m}^*)$; furthermore,
$$d(\mathbf{x}, j) = \Tr_{\mathcal{F}}\big(\rho_{d^*,\vec{m}^*} (n_{\mathbf{x} j}-1)\big), \qquad
\vec{m}(\mathbf{x}, j) = \Tr_{\mathcal{F}}\big(\rho_{d^*,\vec{m}^*} \vec{s}_{\mathbf{x} j}\big),$$
i.e., $d(\mathbf{x}, j) \in [-1,1]$ is the expected density at site $(\mathbf{x},j)$ in the state $\rho_{d^*,\vec{m}^*}$ normalized so that $d(\mathbf{x}, j) = 0$ corresponds to half-filling, and $\vec{m}(\mathbf{x}, j)$ is the expected spin at site $(\mathbf{x},j)$ in the state $\rho_{d^*,\vec{m}^*}$.
Proofs of the above facts can be found in \cite{LLupdate}.

\subsection{Restricted Hartree--Fock theory}\label{rHFapp} 
In restricted Hartree--Fock theory, one minimizes $\Omega$ over only a subset of all quasi-free states. In this paper, we choose to consider states with staggered expected density and staggered expected spin of the form
\begin{align}\label{dvecmdef}
d(\mathbf{x},j)
= d_0 + d_1 (-1)^{x_1 + x_2 + j-1},
\qquad
\vec{m}(\mathbf{x},j) = (m_0 + m_1 (-1)^{x_1 + x_2 + j-1}) \vec{e},
\end{align} 
where $d_0, d_1, m_0, m_1 \in \R$ are constants and $\vec{e} \in \R^3$ is a fixed unit vector. 
By spin rotation invariance, we may, without loss of generality, assume that $\vec{e} = \vec{e}_3 = (0,0,1)$. 
Abusing notation slightly and writing $G(d_0, d_1, m_0, m_1)$ for $G(\rho_{d,\vec{m}})$ and $\rho_{d_0, d_1, m_0, m_1}$ for $\rho_{d,\vec{m}}$, we find (in analogy with the unrestricted case) that if $(d_0^*, d_1^*, m_0^*, m_1^*)\in \R^4$ is a stationary point of $G$ such that
$$G(d_0^*, d_1^*, m_0^*, m_1^*) = \min_{m_0, m_1 \in \R}\max_{d_0,d_1\in \R} G(d_0, d_1, m_0, m_1),$$
then the quasi-free state $\rho_{d_0^*, d_1^*, m_0^*, m_1^*}$ is a minimizer of $\Omega$ among all quasi-free states of the form $\rho_{d_0, d_1, m_0, m_1}$, and $\Omega(\rho_{d_0^*, d_1^*, m_0^*, m_1^*}) = G(d_0^*, d_1^*, m_0^*, m_1^*)$; furthermore,
$$d(\mathbf{x}, j) = \Tr_{\mathcal{F}}\big(\rho_{d_0^*, d_1^*, m_0^*, m_1^*} (n_{\mathbf{x} j}-1)\big) , \qquad
\vec{m}(\mathbf{x}, j) = \Tr_{\mathcal{F}}\big(\rho_{d_0^*, d_1^*, m_0^*, m_1^*} \vec{s}_{\mathbf{x} j}\big).$$

The states (\ref{dvecmdef}) are translation invariant by two states, which means that the term $\Tr_{\mathcal{H}}\big[\Ln_T(h)\big]$ in (\ref{calGdef}) can be computed using Fourier transformation. 
Consider the Fourier transformed basis $\{|\mathbf{k} j \sigma\rangle\}_{\mathbf{k} \in \Lambda_L^*,  j \in \{1,2\}, \sigma \in \{ \uparrow, \downarrow\}}$ for the single-particle Hilbert space $\mathcal{H} \cong \C^{4L^2}$ defined by
\begin{align}\label{kbasisdef}
\langle \mathbf{x} j \sigma |\mathbf{k} j' \sigma' \rangle = \frac{1}{L} e^{i\mathbf{k} \cdot \mathbf{x}} \delta_{j j'} \delta_{\sigma \sigma'}
\end{align}
with the dual lattice $\Lambda_L^*$ given by
$$\Lambda_L^* = \bigg\{(k_1, k_2) \in \frac{2\pi}{L}\Z^2 \,\bigg|\, -\pi \leq k_1 < \pi\;\; \text{and}\; -\pi \leq k_2 < \pi \bigg\}.$$
The linear map $h$ is represented by the following matrix in the basis $\{|\mathbf{k} j \sigma\rangle\}$:
\begin{align*}
h_{\mathbf{k} j \tau, \mathbf{k}' j' \tau'} 
=&\; \delta_{\mathbf{k}\mathbf{k}'} \delta_{\tau \tau'} \begin{pmatrix}
 - 2 t(\cos(k_1) + \cos(k_2)) -\mu& -t_{\perp} \\
-t_{\perp}  &   - 2 t(\cos(k_1) + \cos(k_2)) -\mu
\end{pmatrix} 
	\\
& +\frac{U}{2} \delta_{\mathbf{k}\mathbf{k}'} \big(d_0\delta_{\tau \tau'} - m_0 \vec{e}_3 \cdot \vec{\sigma}_{\tau \tau'}\big) \begin{pmatrix} 	1 & 0
	\\
0 & 1
 \end{pmatrix}
 	\\
& + \frac{U}{2}\delta_{\mathbf{k},\mathbf{k}' + \mathbf{Q}} \big(d_1 \delta_{\tau \tau'} - m_1 \vec{e}_3 \cdot \vec{\sigma}_{\tau \tau'}\big)
 \begin{pmatrix} 1 & 0
	\\
0 & -1
 \end{pmatrix},
\end{align*}
where the rows correspond to $j = 1,2$, the columns correspond to $j' = 1,2$, $\mathbf{Q} \in \R^2$ is the vector $\mathbf{Q} := (\pi, \pi)$, and $\mathbf{k}'+\mathbf{Q}$ is computed modulo $2\pi\Z^2$, so that $\mathbf{k}'+\mathbf{Q} \in \Lambda_L^*$.
We observe that $h$ is block-diagonal in the basis $\{|\mathbf{k} j \sigma\rangle\}$ except that for each $(\mathbf{k}, \sigma)$, the four elements $\mathbf{k}1{\sigma}$, $\mathbf{k} + \mathbf{Q} 1{\sigma}$, $\mathbf{k}2{\sigma}$, and $\mathbf{k} + \mathbf{Q} 2{\sigma}$ are coupled via the $4 \times 4$ matrix
\begin{align*}
& \begin{pmatrix} 
h_{\mathbf{k} 1 \sigma, \mathbf{k} 1 \sigma} & 
h_{\mathbf{k} 1 \sigma, \mathbf{k} + \mathbf{Q} 1 \sigma} &
h_{\mathbf{k} 1 \sigma, \mathbf{k} 2 \sigma} & 
h_{\mathbf{k} 1 \sigma, \mathbf{k} + \mathbf{Q} 2 \sigma} 
	\\
h_{\mathbf{k} + \mathbf{Q}  1 \sigma, \mathbf{k} 1 \sigma} & 
h_{\mathbf{k} + \mathbf{Q}  1 \sigma, \mathbf{k} + \mathbf{Q} 1 \sigma} &
h_{\mathbf{k} + \mathbf{Q}  1 \sigma, \mathbf{k} 2 \sigma} & 
h_{\mathbf{k} + \mathbf{Q}  1 \sigma, \mathbf{k} + \mathbf{Q} 2 \sigma} 
	\\
h_{\mathbf{k} 2 \sigma, \mathbf{k} 1 \sigma} & 
h_{\mathbf{k} 2 \sigma, \mathbf{k} + \mathbf{Q} 1 \sigma} &
h_{\mathbf{k} 2 \sigma, \mathbf{k} 2 \sigma} & 
h_{\mathbf{k} 2 \sigma, \mathbf{k} + \mathbf{Q} 2 \sigma} 
	\\
h_{\mathbf{k} + \mathbf{Q} 2 \sigma, \mathbf{k} 1 \sigma} & 
h_{\mathbf{k} + \mathbf{Q} 2 \sigma, \mathbf{k} + \mathbf{Q} 1 \sigma} &
h_{\mathbf{k} + \mathbf{Q} 2 \sigma, \mathbf{k} 2 \sigma} & 
h_{\mathbf{k} + \mathbf{Q} 2 \sigma, \mathbf{k} + \mathbf{Q} 2 \sigma} 
\end{pmatrix}
	\\
&\hspace{4cm}= \begin{pmatrix} \tilde{h}(d_0, d_1, r m_0, r m_1) & -t_{\perp} I_{2\times 2} \\  -t_{\perp} I_{2\times 2} & \tilde{h}(d_0, -d_1, r m_0, - rm_1) \end{pmatrix},
\end{align*}
where $r = +1$ if $\sigma= \,\uparrow$ and $r = -1$ if $\sigma= \, \downarrow$, $I_{2\times 2}$ is the $2 \times 2$ identity matrix, and
\begin{align*}
& \tilde{h}(d_0, d_1, m_0, m_1) := 
\begin{pmatrix} 
\varepsilon(\mathbf{k}) -\mu + \frac{U(d_0-m_0)}{2} & 
  \frac{U(d_1-m_1)}{2}   
     	\\
\frac{U(d_1-m_1)}{2}  &
\varepsilon(\mathbf{k}+\mathbf{Q}) -\mu + \frac{U(d_0-m_0)}{2} 
 \end{pmatrix}
\end{align*}
with $\varepsilon(\mathbf{k}) := - 2 t(\cos(k_1) + \cos(k_2))$.
Using that $\varepsilon(\mathbf{k}+\mathbf{Q}) = -\varepsilon(\mathbf{k})$, 
we find that this $4 \times 4$ matrix has the four eigenvalues $\{E_{r,r',r''}(\varepsilon(\mathbf{k}))\}_{r',r''=\pm 1}$, where
$$E_{r,r',r''}(\epsilon) := \frac{U}{2}(d_0 - r m_0) - \mu + r' \sqrt{\Big(\frac{U}{2}(d_1 - r m_1)\Big)^2 + (\epsilon + r'' t_{\perp})^2}.$$
Hence
\begin{align*}
\Tr_{\mathcal{H}}\big[\Ln_T(h)\big]
 &= \sum_{\mathbf{k} \in \Lambda_{L, \half}^*} \sum_{r,r',r''=\pm 1} \Ln_T(E_{r,r',r''}(\varepsilon(\mathbf{k}))) 
 	\\ 
 &= \frac{1}{2} \sum_{\mathbf{k} \in \Lambda_L^*} \sum_{r,r',r''=\pm 1} \Ln_T(E_{r,r',r''}(\varepsilon(\mathbf{k}))),
\end{align*}
where 
$$\Lambda_{L, \half}^* := \bigg\{(k_1, k_2) \in \frac{2\pi}{L}\Z^2 \,\bigg|\, -\pi \leq k_1 < \pi\;\; \text{and} \;  -\pi \leq k_2 < 0\bigg\}$$
includes exactly one of the points $\mathbf{k}$ and $\mathbf{k} + \mathbf{Q}$ for each $\mathbf{k} \in \Lambda_L^*$.
Consequently, writing $\mathcal{G}(d, \vec{m}) = \frac{G(d, \vec{m})}{2L^2}$, we obtain
\begin{align*}
\mathcal{G} = \frac{U}{4} \big(m_0^2+ m_1^2 - d_0^2 - d_1^2\big)
- \frac{1}{4L^2} \sum_{\mathbf{k} \in \Lambda_L^*} \sum_{r,r',r''=\pm 1} \Ln_T(E_{r,r',r''}(\varepsilon(\mathbf{k}))), 
\end{align*}
which in the thermodynamic limit $L \to \infty$ becomes
\begin{align*} 
\mathcal{G} 
 & = \frac{U}{4} \big(m_0^2+ m_1^2 - d_0^2 - d_1^2\big)
- \frac{1}{4} \int_{[-\pi, \pi]^2} \sum_{r,r',r''=\pm 1} \Ln_T(E_{r,r',r''}(\varepsilon(\mathbf{k})))  \frac{d^2\mathbf{k}}{(2\pi)^2} 
	\\ 
& = \frac{U}{4} \big(m_0^2+ m_1^2 - d_0^2 - d_1^2\big)
- \frac{1}{4} \int_\R N_0(\epsilon) \sum_{r,r',r''=\pm 1} \Ln_T(E_{r,r',r''}(\epsilon)) d\epsilon, 
\end{align*}
where $N_0(\epsilon)$ is given in (\ref{2Ddensityofstates}), and we have assumed that the functions are sufficiently smooth for the Riemann sum approximation $\frac{1}{L^2} \sum_{\mathbf{k} \in \Lambda_{L}^*} \to \int_{[-\pi,\pi]^2}  \frac{d\mathbf{k}}{(2\pi)^2}$ to apply.
One can write this as 
\begin{align}\label{cG}  
\mathcal{G} = \frac{U}{4} \big(m_0^2+ m_1^2 - d_0^2 - d_1^2\big) - \frac{1}{2} \int_\R N_{t_\perp}(\epsilon) \sum_{r,r'=\pm 1} \Ln_T(E_{r,r'}(\epsilon)) d\epsilon 
\end{align} 
with $N_{t_\perp}(\epsilon)$ in (\ref{bilayerdensityofstates}) and  $E_{r,r'}(\epsilon):=E_{r,r',0}(\epsilon)$ equal to 
\begin{align}\label{Err}  
E_{r,r'}(\epsilon) := \frac{U}{2}(d_0 - r m_0) - \mu + r' \sqrt{\Big(\frac{U}{2}(d_1 - r m_1)\Big)^2 + \epsilon^2}.
\end{align}

\subsection{The case of half-filling}
The bilayer lattice $\Lambda \times \{1,2\}$ is bipartite in the sense that it is the disjoint union of two sublattices $A, B \subset \Lambda_L \times \{1,2\}$ such that $t_{\mathbf{x}\mathbf{y}} = 0$ whenever $\mathbf{x}, \mathbf{y} \in A$ or $\mathbf{x},\mathbf{y} \in B$.
Moreover, the hopping matrix is invariant under translations in the $x$ and $y$ directions, and also under the map that interchanges the two layers. 
Thus, if $\rho_{d^*,\vec{m}^*}$ is a minimizer of $\Omega$ among all quasi-free states for $\mu = 0$, then
\begin{enumerate}
\item by \cite[p. 12]{BP1996}, $d^* \equiv 0$ (i.e., the system is at half-filling) and
\item by \cite[Theorem 4 and Theorem 5]{BP1996}, $\vec{m}^*(\mathbf{x},j) = m_1 (-1)^{x_1 + x_2 + j-1} \vec{e}$ for some constant $m_1 \in \R$ and some fixed unit vector $\vec{e} \in \R^3$.
\end{enumerate}
This means that, for $\mu = 0$, our assumption that the variational fields $d$ and $\vec{m}$ have the form in (\ref{dvecmdef}) is without loss of generality, and, moreover, it holds that $d_0 = d_1 = m_0 = 0$.
Assuming that $\mu = 0$ and setting $d_0 = d_1 = m_0 = 0$ and $\Delta_{\AF} = \frac{U}{2} m_1$, we find
\begin{align} \label{mathcalGAF}
\mathcal{G} & = \frac{\Delta_{\AF}^2}{U} 
- 2\int_\R N_{t_\perp}(\epsilon) \Ln_T(\sqrt{\Delta_{\AF}^2 + \epsilon^2}) d\epsilon 
\end{align}
and the mean-field equation $\frac{\partial \mathcal{G}}{\partial m_1}  = 0$ reduces to
\begin{align}
\frac{\Delta_{\AF}}{U} 
= \int_\R N_{t_\perp}(\epsilon) \Ln_T'(\sqrt{\Delta_{\AF}^2 + \epsilon^2}) \frac{\Delta_{\AF}}{\sqrt{\Delta_{\AF}^2 + \epsilon^2}}d\epsilon.
\end{align}
Dividing by $\Delta_{\AF}$ and inserting $\Ln_T'(E) = \frac{1}{2} \tanh \left(\frac{E}{2T}\right)$ yields equation (\ref{DeltaAFeq}).

\subsection{Phase diagrams away from half-filling}\label{awayfromhalffillingapp}
As explained in Appendix~\ref{rHFapp}, the restricted Hartree--Fock states at fixed values of $T\geq 0$ and $\mu\in\R$ are determined by  
\begin{align} 
\mathcal{G}(d_0^*, d_1^*, m_0^*, m_1^*) = \min_{m_0, m_1 \in \R}\max_{d_0,d_1\in \R} \mathcal{G}(d_0, d_1, m_0, m_1)
\end{align} 
with $\mathcal{G}(d_0, d_1, m_0, m_1)$ in \eqref{cG}--\eqref{Err} and, at such a solution $(d_0^*, d_1^*, m_0^*, m_1^*)$, doping is given by $\nu=d_0^*$. 

We found in extensive numerical experiments that only solutions of the form \eqref{MF} occur. 
It would be interesting to prove this by mathematical arguments.

\bigskip\noindent {\bf Acknowledgements.} 
The authors are grateful to Immanuel Bloch for helpful discussions and for bringing the significance of the bilayer Hubbard model to their attention.
They also would like to thank Joscha Henheik, Asbj{\o}rn B{\ae}kgaard Lauritsen and Valdemar Melin for collaboration on closely related projects.
E.L. acknowledges support from the Swedish Research Council, Grant No. 2023-04726. 
J.L. acknowledges support from the Swedish Research Council, Grant No.\ 2021-03877.

\bigskip\noindent {\bf Conflict of interest.} 
Both authors declare that they have no conflicts of interest.

\medskip\noindent {\bf Data Availability.} 
The datasets generated to produce the figures in this study are available upon reasonable request.

\medskip\noindent {\bf Generative AI Declaration.} 
No generative AI tools have been used to produce the mathematical content or the presentation of this manuscript.

\bibliographystyle{plain}

\end{document}